# Fundamental Limits of Detection in the Far Infrared

S.P. Denny[1], J. Y. Suen[2], P. M. Lubin[1]

1. Department of Physics, University of California, Santa Barbara, California 93106, USA

2. Department of Electrical and Computer Engineering, University of California, Santa Barbara, California 93106, USA

**Corresponding Author**: P. M. Lubin, lubin@deepspace.ucsb.edu. Phone: +1-805-893-8432, Fax: +1-805-893-8432

## Abstract

We study the fundamental limits of detection for astrophysical observations in the far infrared. Understanding these fundamental limits is critical to the planning and analysis of experiments in this region. We specifically characterize the difficulties associated with observing in the 0.1-10 THz (30-3000 µm) regime including extraterrestrial, atmospheric, and optical emission. We present signal, noise, and integration time models for selected terrestrial, aircraft, balloon, and space missions. While ground based telescopes offer the great advantage of aperture size, and hence angular resolution, they suffer from the relatively low transmission and high radiance of the atmosphere, particularly for wavelengths less than 500 µm. Space telescopes are the inverse; they are limited by a small aperture, while an airborne telescope is constrained by both. Balloon-borne telescopes provide an option over much of the band. A quantitative understanding of this is critical in comparing the sensitivity of various experiments and in planning the next generation of missions. As representative sources we use the luminous far-IR dusty galaxies NGC 958 and Mrk 231, but the same formalism can be applied to any source. In this paper we focus on continuum emission but a future paper will focus on line emission.

## 1. Introduction

Our galaxy is a typical spiral galaxy in the visible bands and as such has been used as a starting model for understanding how to optimally study high redshift galaxies. However, while this appears to be valid in the visible and near IR it may not be so in the far-IR where the fraction of radiation due to dust emission can vary greatly. Galaxies are basically a collection of gravitationally bound stars embedded in a "soup" of gas, dust, and dark matter. The earliest galaxies are likely to have a significantly different amount of gas and dust compared to later evolved galaxies. The emission of galaxies can be largely traced to the "fusion" component of stars that is dominated by sources that are thousands of degrees Kelvin, while the dust that is largely heated by starlight is typically tens of Kelvins, though multiple dust temperatures are also seen. Thus there are two primary emission regions. One is associated with the visible and near IR peaking near 1 µm and the other associated with the relatively cold dust peaking near 100 µm. The presence of both simple and complex molecular gases can also lead to a number of emission and absorption lines that are critical to the overall formation of the galaxy as they provide a means to radiate away significant energy and momentum, without which gravitation collapse to the large densities ultimately needed for star formation would not be possible. These absorption and emission lines provide a damping mechanism and are critical to help us understand the formation of galactic and stellar structures.

We do not yet know the age, and hence redshift, of the first generation of galaxies but we expect it will be around redshift 10, or some few hundred million years after the "Big Bang". Finding these earliest galaxies and tracing their formation and evolution is a high priority for modern cosmology and will help us better understand the roles of dark matter and dark energy in the early universe. Naively thinking of galaxies as having two emission mechanisms, stellar emission processes near 1 µm and reradiated emission of absorbed starlight from dust emission near 100 µm, we would conclude that at redshift 10 we would want to observe near 10 and 1000 µm. Indeed in a simplistic sense this is what we do want to do, but we can learn more if we study the ranges from 1-30 µm and 30-3000 µm and measure the detailed spectral energy distribution (SED) as well as the various emission and absorption lines. The region from 1-30 µm is very difficult to observe from the ground due to the extremely high emission from our atmosphere and from warm optics that is typically compatible with ground-based measurements. The JWST mission will be extremely sensitive in the 1-30 µm band where a space-based mission is critical. Detailed mission comparisons will be discussed in a later paper. The region from 30-1000 µm is also quite difficult to observe from the ground due to absorption by our atmosphere, in particular the critical region near 100 µm is nearly opaque. In this paper we focus on the fundamental limits from 30 to 3000 µm (0.1 - 10 THz) where the redshifted dust emission and far-IR cooling lines can be observed. As a rule of thumb, roughly equal power is emitted by a galaxy from the visible stellar (fusion) component as from the reradiated dust component. We are used to seeing incredible images from the stellar component but much less so from the dust component due to technical maturity and the great difficulty of seeing through our terrestrial atmosphere. Recent missions such as the Planck and Herschel show the beautiful complexity of dust in our own and other galaxies.

In this paper, we set out to model the fundamental limits of observing high redshift galaxies from their dust emission. The detection of a source can be separated into noise and signal. In the first section of this paper, we discuss the various sources of noise. For these observations, the CMB, cosmic infrared background, zodiacal, galactic, mirror, and atmospheric emissions are significant.

Next, we discuss the signal emitted from two representative dusty galaxies, specifically the spectral energy distribution (SED). For observations within the Earth's atmosphere we model the highly complex frequency-dependent atmospheric attenuation and emission, all relevant foregrounds, as well optical emission. We then compute the total power received for a given system. Finally, we combine the effects of signal and noise by calculating the integration time for a selection of ground, aircraft, balloon-borne and space based systems to achieve a given Signal to Noise Ratio (SNR).

## 1.2 Representative Telescopes

As a background to our modeling, we present five general types of systems, which are representative of operating or proposed programs. They are ground based, warm optics balloon-borne, cold optics balloon-borne, airborne, and space-based systems. The purpose of this paper is to explore fundamental observation limits, and so we will generally assume ideal systems; 100% aperture efficiency, 100% QE detectors and no optical emission other than the primary mirror except in cases of existing or fully proposed instruments. These idealized assumptions place each system on equal "footing". We note optical efficiency, emissivity and detector efficiency can be far from ideal. Starting from idealized assumption allows us to quickly add real system parameters in future work.

As a representative ground based system, we consider a 25 m telescope, located at Cerro Chajnantor, Chile. This resembles the proposed Cerro Chajnantor Atacama Telescope (CCAT). The Cerro Chajnantor site is representative of a very good ground based high-altitude location. The choice of location for far-IR ground based telescopes is complicated by competing factors where access and operational costs become extremely important. The dominating issue for performance here is primarily water vapor content (precipitable water vapor or PWV, with units of $g/cm^2$) and not altitude. Here altitude is usually a means to an end (low PWV) and not an end in itself.

As an airborne model, we studied the sensitivity of the Stratospheric Observatory for Infrared Astronomy (SOFIA) a 2.5 m telescope flown on a modified 747 aircraft that flies at altitudes up to about 14 km. Its optics run relatively warm, typically 230 K, matching the air temperature outside the aircraft. It also has relatively high emissivity in the far-IR. Out of the telescopes modeled, SOFIA is the only system that has been constructed. As built, SOFIA uses an un-cooled and relatively emissive optical chain, which we modeled as an equivalently higher mirror emissivity. We compared both an idealized airborne observatory with much lower emissivity and the same system with the much higher design goal, though still optimistic, emissivity.

We also considered telescopes flown on high-altitude balloons. The balloon-borne systems were assumed to have either a 2.2 m or a 10 m diameter mirror. Getting a 10 m telescope to 40 km is on the edge of possibility for current technology, and so for this case we limit the temperature of the mirror to 230 K, which is roughly the ambient temperature at that altitude. A 2.2 m diameter is a much more reasonable proposition (we have already developed and flown multiple telescopes of this size), and so we assume that it can be run at any temperature greater than or equal to 2K. For the purposes of comparing these cases, we present data for a balloon-borne 2.2 m telescope operating at 2 K and 230 K as well a balloon-borne 10 m telescope operating at 230K all observing at an altitude of 40 km.

Finally we consider a far-IR space based telescope. The proposed Space Infrared Telescope for Cosmology and Astrophysics (SPICA) served as a template for an orbital mission with optics cooled to 4.5

K and a 3.5 m aperture (Nakagawa et al. 1998). For simplicity, we refer to the template as SPICA in later sections.

Finally we note that we have considered literally hundreds of models and present only a few representative ones here. Readers with special interests are encouraged to contact us. We note the science case for observing in the far-IR is eloquently presented in the NAS 2010 Decadal survey (New Worlds, New Horizons - Committee for a Decadal Survey of Astronomy and Astrophysics, 2010).

## 2. Noise in the THz regime

### 2.1 Basic quantities

We first set out the fundamental description of noise that we will use. Similar to how electrons inside a conductor will undergo thermal excitation and generate Johnson-Nyquist noise, any noise source can be described as having an equivalent antenna temperature $T_a$, where the total noise power, $P_{T_a}(\nu, \Delta\nu)$ [watts], is

$$P_{T_a}(\nu, \Delta\nu) = k_b \Delta\nu \, m \, T_a \, , \qquad \text{(eqn 2.1)}$$

for a bandwidth $\Delta\nu$ and Boltzmann's constant $k_b$. This is the limit in the long wavelength (Rayleigh-Jeans - RJ) regime for a blackbody whose physical temperature T is equal to the antenna temperature $T_a$ . The variable $m$ represents the number of modes accepted by the detectors. Here we assume that the detector accept unpolarized radiation to allow for greatest sensitivity, hence $m = 2$. Since each of the noise sources are uncorrelated the total system antenna temperature (power) is simply the sum of each component in the limit of minimal absorption of any source,

$$T_{sys} = \sum_i T_{a_i}. \qquad \text{(eqn 2.2)}$$

In the case of transmission through the Earth's atmosphere we first reduce each extraterrestrial component by the atmospheric transmission. It is important to note that antenna temperature is not equivalent to physical temperature. A blackbody at physical temperature $T_b$ has an equivalent antenna temperature $T_a$ at frequency $\nu$ of,

$$T_a = \frac{T_b x}{e^x - 1} \quad \text{where} \quad x = \frac{h\nu}{k_b T_b}. \qquad \text{(eqn 2.3)}$$

For an optical system of collecting area $A$, a solid angle field of view $\Omega$, and observation bandwidth $\Delta\nu$, and intensity $I(\nu)$ $[W/(m\hat{}2 \, srHz)]$ the measured power will be,

$$P(\nu, \Delta\nu) = I(\nu) A \Omega \Delta\nu. \qquad \text{(eqn 2.4)}$$

For convenience, we define the power per unit bandwidth for a system $\boldsymbol{s}(\nu)$ $[W/Hz]$ where:

$$\boldsymbol{s}(\nu) = I(\nu) A \Omega \; = m k_b T_a \qquad \text{(eqn 2.5)}$$

Combining equations 2.1, 2.4 and 2.5

$$T_a(\nu) = \frac{P(\nu,\Delta\nu)}{2k_b\Delta\nu} = \frac{I(\nu)A\Omega}{2k_b} = \frac{s(\nu)}{2k_b}. \quad \text{(eqn 2.6)}$$

If we assume diffraction limited optics,

$$A\Omega = \lambda^2 = \frac{c^2}{\nu^2}, \quad \text{(eqn 2.7)}$$

which can be combined with equation 2.6 for the result

$$T_a(\nu) = \frac{I(\nu)c^2}{2k_b\nu^2} = \frac{I(\nu)\lambda^2}{2k_b} \quad \text{(eqn 2.8)}$$

Equation 2.8 calculates antenna temperature given a frequency-dependent intensity (assuming dual polarization) profile, while equation 2.3 does the same given a blackbody temperature profile.

## 2.2 Background Limiting Noise Generated (BLING)

Antenna temperature is a traditional and concise way to express the noise generated in a system, but it is preferable to express noise in a fashion that allows for easy comparison with the desired signal received from an astronomical source. The quantity that is commonly used for detectors is the noise equivalent power (NEP), which is the signal power necessary such that the overall signal-to-noise ratio is unity $(SNR = 1)$ in one second of observation. For detectors that are limited by photon statistics, the background limited infrared photometry (BLIP) limit applies. Both of these quantities are in units of $W/\sqrt{Hz}$ and directly relate noise power with post detection bandwidth and integration time. These quantities can be compared with the flux of a source, which is measured in units of $W/(m^2Hz)$. We note that the reference to "Hz " in the source flux relates to the pre-detection bandwidth while the “Hz” in the NEP and BLIP (and BLING below) refer to the post-detection bandwidth. These are easily confused and care should be taken to make sure that they are not.

In the interest of detailing the noise of our systems in a uniform and useful way, we define a quantity called the "Background Limiting Noise Generated" (BLING) for each noise source, with the same $W/\sqrt{Hz}$ units of NEP. This is often referred to as the "BLIP Limit". We generalize this to each source we consider as well as to the combination of all external and internal sources. An example of such a source is atmospheric radiance or galactic emissions, which are external to the telescope but are intrinsic to any observation made. These emissive and absorptive sources impair the detection of the desired signal and therefore we can consider them as components of the overall background of the telescope system.

The BLING, $\zeta$, is simply,

$$\zeta = P_N t^{1/2}, \quad (2.9)$$

given a noise power $P_N$ $[watts]$ and integration time $t$. We can now show the relationship between BLING and antenna temperature. The power corresponding to a certain antenna temperature over a bandwidth $\Delta\nu$, centered around $\nu_0$ can be found by integrating equation 2.5

$$P(\nu_0, \Delta\nu) = \int_{\nu_0-\frac{\Delta\nu}{2}}^{\nu_0+\frac{\Delta\nu}{2}} \boldsymbol{s}(\nu)d\nu. \quad (2.10)$$

Since the energy of a photon, $E_{photon} = h\nu$, the total photon flux $F$ in number of photons with a certain energy per unit time is

$$F(\nu_0, \Delta\nu) = \frac{P(\nu_0,\Delta\nu)}{E_{photon}(\nu_0,\Delta\nu)} = \int_{\nu_0-\frac{\Delta\nu}{2}}^{\nu_0+\frac{\Delta\nu}{2}} \frac{\boldsymbol{s}(\nu)}{h\nu} d\nu \quad (2.11)$$

The actual number of photons collected in some integration time $t$ is simply $N(\nu_0, \Delta\nu) = F(\nu_0, \Delta\nu)t$. The uncertainty, or standard deviation, in the number of incoming photons in time $t$ is then

$$\delta N(\nu_0) = N(\nu_0)^{\frac{1}{2}} = (F(\nu_0)t)^{\frac{1}{2}} = \left(t \int_{\nu_0-\frac{\Delta\nu}{2}}^{\nu_0+\frac{\Delta\nu}{2}} \frac{\boldsymbol{s}(\nu)}{h\nu} d\nu\right)^{\frac{1}{2}}, \quad (2.12)$$

and the BLING is the product of the uncertainty of photons and the energy of the photons.

$$\zeta(\nu_0) = E_{photon}\delta N(\nu_0) = \left(t \int_{\nu_0-\frac{\Delta\nu}{2}}^{\nu_0+\frac{\Delta\nu}{2}} h\nu\boldsymbol{s}(\nu)d\nu\right)^{\frac{1}{2}}, \quad (2.13)$$

which is in units of $W/\sqrt{Hz}$ (here we set $t = 1\ s$). Finally, it is useful to define the detection bandwidth as a fractional bandwidth, that is $R = \frac{\nu}{\Delta\nu}$, so

$$\zeta(\nu_0) = \left(2 \int_{\nu_0-\frac{\nu_0}{2R}}^{\nu_0+\frac{\nu_0}{2R}} hk_B\nu T_a(\nu)d\nu\right)^{\frac{1}{2}} = \left(\int_{\nu_0-\frac{\Delta\nu}{2R}}^{\nu_0+\frac{\nu_0}{2R}} h\nu\boldsymbol{s}(\nu)d\nu\right)^{\frac{1}{2}}. \quad (2.14)$$

This equation relates antenna temperature to BLING for arbitrary frequency-dependent antenna temperatures. Here we assume dual polarization. There is a technical difference between $W/\sqrt{Hz}$ and $W\sqrt{s}$ (normally $W\sqrt{s} = \sqrt{1/2}W/\sqrt{Hz}$, but this depends on the roll off of the post detection filter used). Here we treat them as the same for simplicity.

Note in all of the following plots we use R=1000 for the spectral resolution unless otherwise stated. This allows us to show most of the atmospheric lines and is appropriate for a spectroscopy mission but a broadband photometry mission will generally have a much smaller R (typically 3-10) and we briefly discuss this at the end of the paper in section 4.2.

| | |
|---|---|
| Fractional Bandwidth $R$ | 1000 |
| Observation angle | 45 degrees from Zenith |
| PWV | Best Quartile (Table 3.2) for ground based observations |
| SED Template | NGC 958, Mrk 231 (Table 3.1) |
| Cosmological matter density $\Omega_M$ | 0.3 |
| Cosmological constant $\Omega_L$ | 0.7 |
| Dark Energy equation of state - w | -1 |
| Cosmological expansion rate $H_0$ (km/s-Mpc) | 70 |
| Antenna Modes | 2 (Diffraction limited, both polarizations) |
| Galactic Latitude and Longitude | Galactic North Pole |
| Ecliptic Latitude | 90 degrees |
| Ecliptic Longitude | 0 degrees |
| Signal to Noise Ratio desired | 5 |

**Table 2.1**: Model parameters used, unless otherwise specified.

## 2.3 Sources of Noise

### 2.3.1 Extraterrestrial Sources

Extraterrestrial sources we modeled were the Cosmic Infrared Background (CIB), the Cosmic Microwave Background (CMB), galactic emission, and zodiacal emission. Each of these sources physically lies between the detector and the target galaxy. We treated these sources as a background or foreground (as appropriate) noise source and we assume they are all diffuse. However we note that while the CIB and CMB are essentially isotropic, galactic and zodiacal emissions are not, and at high enough angular resolution much of the CIB may be resolved. The results of all four of these sources are shown in Figure 2.1.

#### 2.3.1.1 Cosmic Infrared Background

The Cosmic Infrared Background (CIB) is of significant concern in the frequency range from 0.1 to 10 THz (30 to 3000 μm) and is thought to be the accumulated light of a large number of distant and dusty sources (Hauser and Dwek 2001, Franceschini et al. 2001). The CIB has a second optical band from 1-10 μm with differing characteristics which are not of direct interest in this paper and will be covered in a future paper. The astrophysical mechanism primarily responsible for photon production in the submillimeter through far-IR waveband is the emission of dust in the galactic interstellar medium (ISM). Stars largely produce visible and UV emission, which is then absorbed by the dust clouds surrounding them. The dust particles are typically heated to tens of degrees Kelvin and then re-radiate photons, mainly with wavelengths between 10 to 1000 μm. (Franceschini et al. 2008). Observations in the far-IR by the Infrared Astronomical Satellite (IRAS) (Soifer et al. 1987) and many others since have helped resolve the source distribution showing that luminous and ultra-luminous IR galaxies (LIRGs and ULIRGs) are relatively rare at low redshifts and increase at higher redshifts, but it is not clear that they dominate at any redshift. IR galaxies are normally associated with intense star formation. Galaxies that emit in the UV on the other hand are typically of lower mass and do not contain much of an ISM with enough dust to be optically thick, and therefore are faint in the IR but can be visible in UV.

Since IRAS, many observatories have added to the observational database of the universe in the IR domain. The major contributors have been two space observatories, the Infrared Space Observatory (ISO) and the Spitzer Space Telescope, Herschel, as well as several ground based telescopes (SCUBA on JCMT (Blain et al. 2002); BOLOCAM and MAMBO on IRAM (Bertoldi et al. 2003)). We used the Franceschini et al. (2008) model of the far-IR CIB which used a multi-wavelength reference model for galaxy evolution. Figure 2.1 shows the modeled CIB peaking in the 100-300 µm range (1-3 THz), with a value of $\sim 10^{-20} W/\sqrt{Hz}$ for a spectral resolution of 1000. At frequencies less than ~0.3 THz, the CIB is not well understood, so we cannot assume any behavior. The current thought is that with sufficient resolution most sources of the CIB can be resolved. ISO was able to resolve about 3-10% of the total CIB as discrete sources while Spitzer measurements detected about 30% of the CIB as discrete sources. With the Herschel space telescope this number is rising to much higher levels with Great Observatories Origins Deep Survey (GOODS) observations nearing 90% identification of discrete CIB sources in their fields (Lagache et al. 2003, Magnelli et al. 2013).

#### 2.3.1.2 Cosmic Microwave Background

The second background we considered was the Cosmic Microwave Background (CMB) from the time of "last scattering" or so-called "recombination". The temperature of the universe was approximately 3,000 K at the time of recombination. The expansion of the universe redshifted these photons to the current value of 2.73 K and provide a blackbody background that is isotropic to 1 part in $10^5$. Because the RMS variations are $\sim 30$ µK (Wright 2003), we used an isotropic model.

The Planck distribution describes this background well. The noise contribution at low frequencies is quite significant, $(\sim 10^{-18}\ W/\sqrt{Hz})$ for R=1000, due to the peak of the blackbody at ~160 GHz. At much higher frequencies, the Planck distribution rapidly rolls off, and as a result the CMB contributes little to cumulative noise.

#### 2.3.1.3 Galactic Emission

Finkbeiner et al. (1999) used IRAS and Cosmic Background Explorer (COBE) data to extrapolate full sky maps of submillimeter and microwave emission from the diffuse interstellar dust in our own galaxy. Using a two component emissivity model, they found excellent agreement with the data from the COBE Far Infrared Absolute Spectrophotometer (FIRAS) instrument.

At lower frequencies de Oliveira-Costa et al. (2008) presented a global sky model for diffuse galactic radio emission to model foreground contamination of cosmological measurements. The source of this emission is dominated by synchrotron radiation at low frequencies (<100 GHz); radiation from ultra-relativistic charged particles being accelerated through a magnetic field. They found a three component model that closely resembles a power law which is a good match to data sets of large-area radio surveys.

Using these two models, we modeled the power spectrum of galactic emission over our entire region of interest for any point in the galaxy. It is important to understand that, unlike the CIB and CMB, galactic emission is highly anisotropic. Observations perpendicular to the galactic plane have far less galactic emission than those along the plane of the disk, particularly towards the center of the galaxy.

The two extremes were the galactic center and the Galactic North Pole (GNP). The difference in BLING is approximately two orders of magnitude higher towards the galactic center. Since we are interested in detection limits, we have used the GNP in the noise and integration time calculations presented later in this paper.

### 2.3.1.4 Zodiacal Dust Emission

The interplanetary dust cloud that is present within our solar system is the final extraterrestrial source of noise we considered. The scattered light and thermal emission of the interplanetary dust (IPD) particles compose the zodiacal dust emission. We use a model that was developed by Kelsall et. al 1998. (1998) to predict and extract the zodiacal emission from the data collected by the COBE Diffuse Infrared Background Experiment (DIRBE).

This model has the capability of modeling zodiacal emission over the DIRBE wavelength bands, which range from 1.25 μm to 240 μm, as well at various dates and ecliptic coordinates. Using this model, we generated zodiacal emission spectra for a variety of ecliptic coordinates and points along the Earth's orbit. Since the goal of this paper is to illustrate the fundamental limits on observing an IR galaxy, we used ecliptic coordinates ecliptic latitude = 90°, ecliptic longitude = 0°. These ecliptic coordinates correspond to the ecliptic north pole, and yield the least zodiacal emission. Though the model depends on time of year, there were minimal temporal variations in intensity while observing towards the ecliptic north pole. In this paper we do not consider missions outside the solar system, which would have vastly less Zodiacal emission. Backscattered sunlight off the IPD is important in the visible, near and mid IR, but much less so in the far-IR.

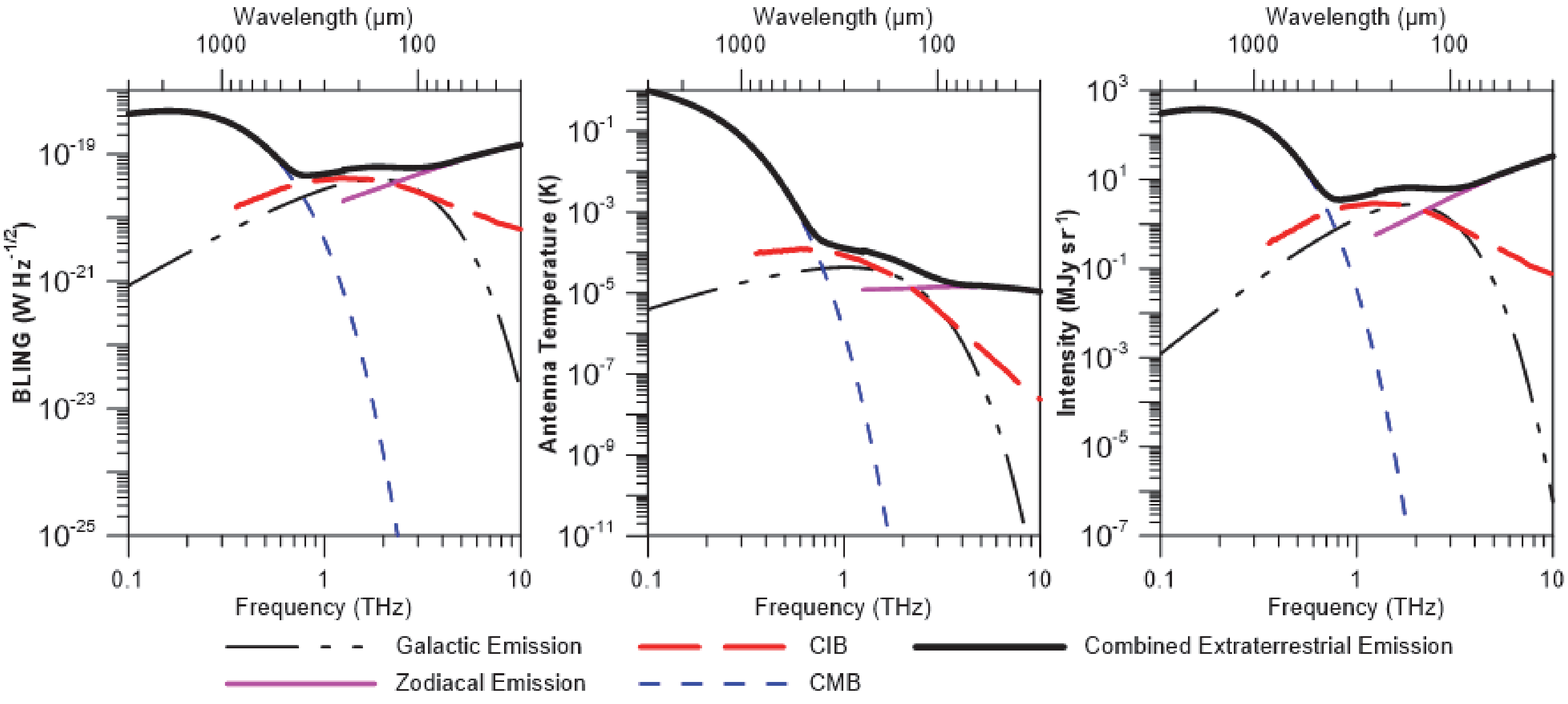

**Figure 2.1**: BLING, antenna temperature and intensity for the four extraterrestrial sources, as well as the cumulative spectrum. For low frequencies (0.1-0.6 THz) the CMB is the dominant contributor. For mid-range frequencies (0.6-3 THz) the CIB and galactic emission are dominant. At high frequencies (3-10 THz) the zodiacal dust emission is dominant. Model parameters are shown in Table 2.1.

### 2.3.2 Thermal Mirror Emission

The mirror of a telescope is a source of photon noise due to its own thermal emission. We assume that the entire mirror is heated uniformly to some operating temperature, $T_0$. If we take the mirror to be a perfect blackbody, it emits according to the Planck distribution with temperature $T_0$. However, a mirror is certainly not a blackbody; to model its emission spectrum we first determine its emissivity.

For simplicity we assume that the mirror is made of pure Aluminum and approximate the emissivity function of the mirror using the Hagen-Rubens formula (Xu et al. 1996).

$$\varepsilon = \left(\frac{16\pi c\epsilon_0}{\lambda\sigma}\right)^{\frac{1}{2}} \qquad \text{(eqn 2.15)}$$

$\lambda$ is wavelength, $\sigma$ is the surface electrical conductivity of Aluminum ($3.538 \times 10^7 \Omega^{-1} m^{-1}$) , $\epsilon_0$ is the permittivity of free space, $c$ is the speed of light (Fig. 2.2). The thermal radiation power generated by a non-blackbody (greybody) surface, $P_{gb}$, at a particular frequency is given by

$$P_{gb} = \varepsilon P_{bb} \qquad \text{(eqn 2.16)}$$

where $P_{bb}$ is the equivalent blackbody power radiated into the system .

If we go back to equation 2.1 and substitute, we find

$$P_{gb} = \varepsilon 2 k_B T_a \Delta\nu = 2 k_B T_{eff} \Delta\nu \qquad \text{(eqn 2.17)}$$

where $T_{eff} = \varepsilon T_a$ is the effective antenna temperature of the mirror surface. $T_{eff}$ is the value for expressing the noise generated by the mirror's thermal emission. Using equation 2.3 we find $T_{eff}$ for a given physical temperature $T_0$ and parameter $x$ as in equation (2.3),

$$T_{eff}(\nu) = \varepsilon(\nu) \frac{x}{(e^x - 1)} T_0. \qquad \text{(eqn 2.18)}$$

Using this equation, we calculated effective antenna temperature and BLING spectra for a variety of mirror operating temperatures. The results are shown in Figure 2.3. In order to significantly reduce BLING in the far IR bands, the mirror should be cooled to cryogenic for maximum sensitivity in the far-IR. In practice, this is normally extremely difficult.

As previously mentioned, the constructed SOFIA telescope uses a significant amount of uncooled optics and has relatively large spillover into the warm aircraft telescope cavity .These significantly increase the overall emissivity and hence were modeled as an equivalent mirror emissivity $\varepsilon = 0.1$ (Becklin, 1997), representing an average value for the measured system optics emissivity. $\varepsilon = 0.1$ is probably an optimistic value in the far IR but we will use it anyway. In addition we also model an idealized SOFIA with total emissivity of a single aluminum mirror with much lower emissivity than the actual telescope. Such a system would be a redesign with cold secondary mirrors.

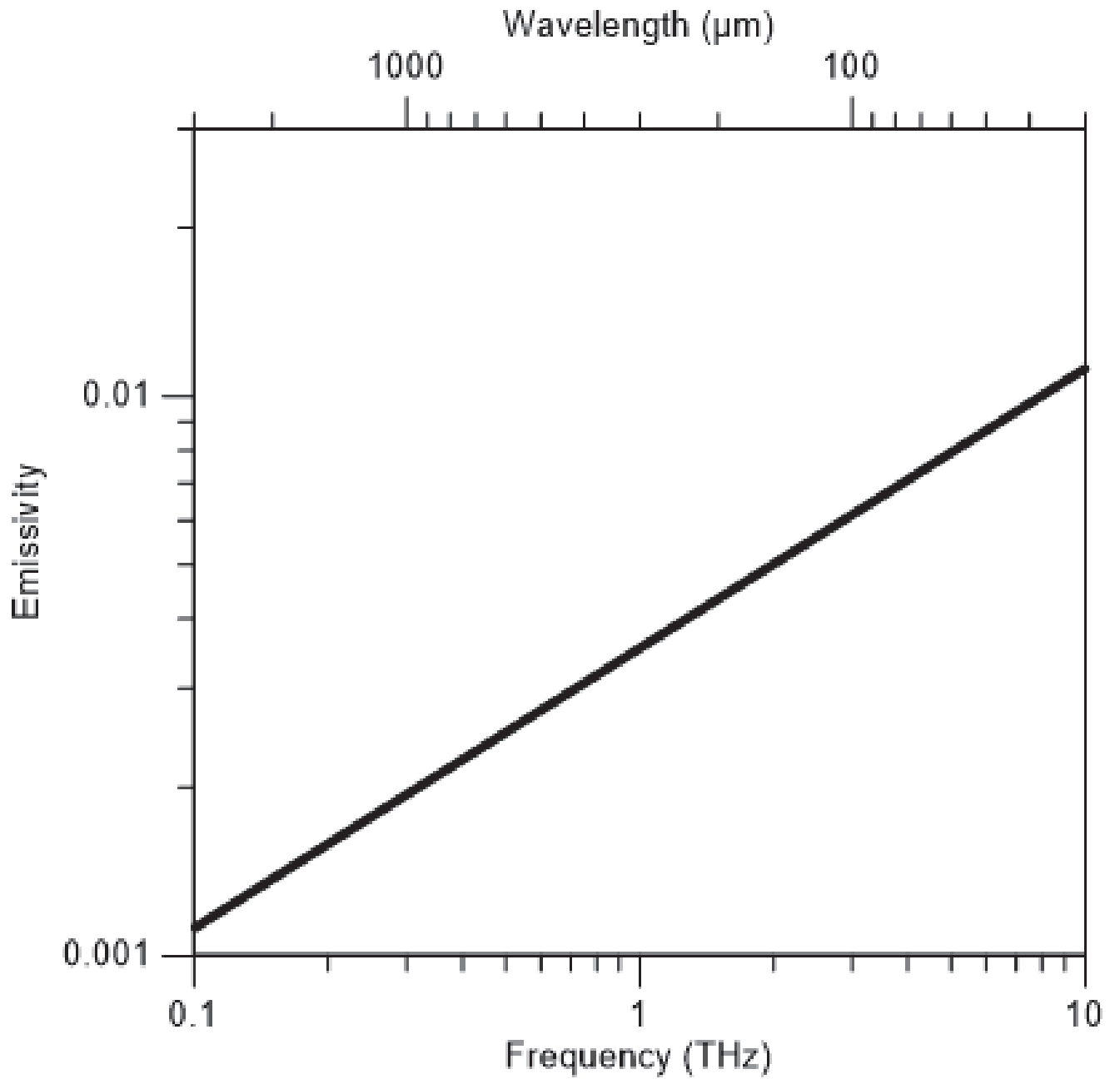

**Figure 2.2**: Modeled emissivity of an aluminum mirror vs frequency. Measured emissivity is comparable for bulk metal polished surfaces. Thin gold on copper is slightly better and sometimes used.

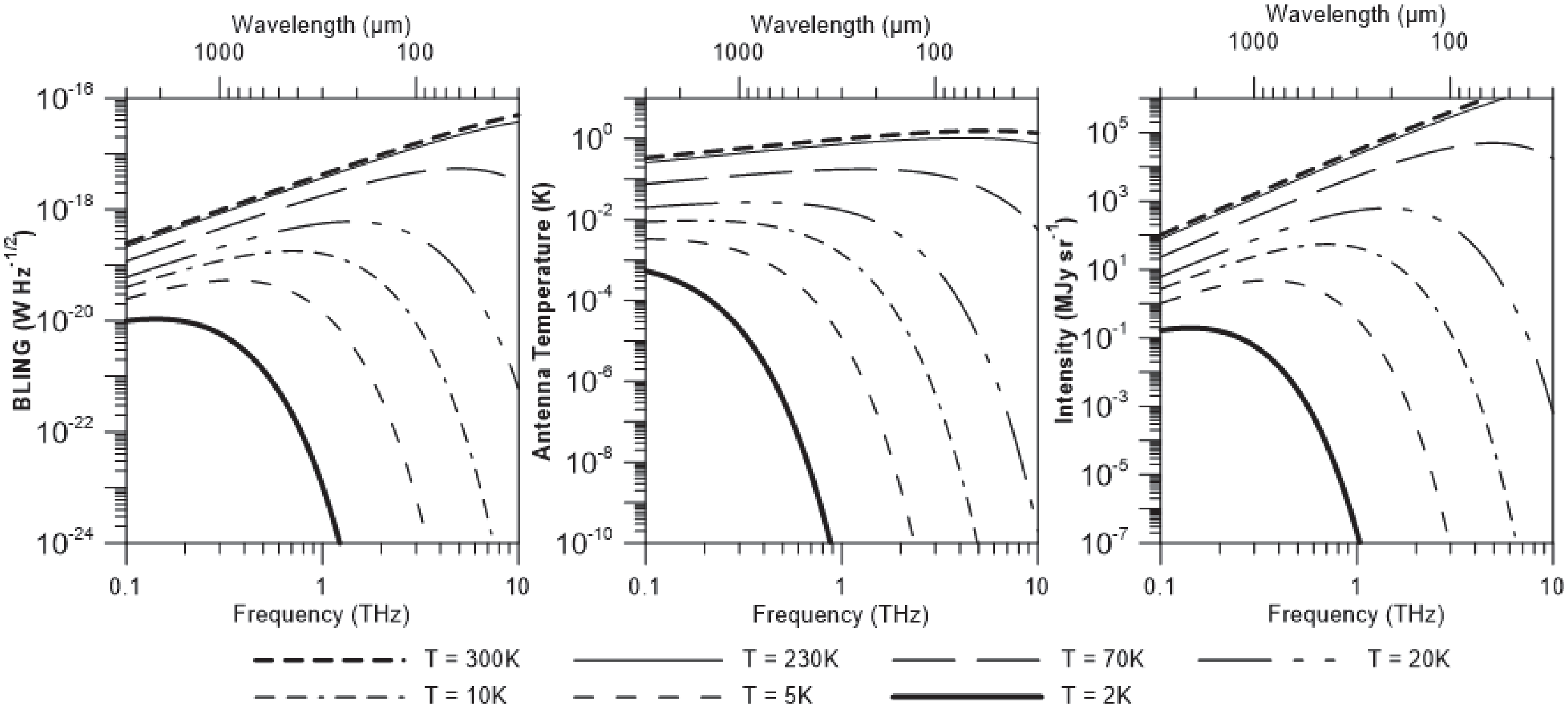

**Figure 2.3:** BLING, antenna temperature and intensity for aluminum mirrors of various temperatures. Note that the total telescope emissivity may not be dominated by mirror emissivity alone, as is the case for some ground based instruments and for SOFIA, which can have significant spillover.

### 2.3.3 Atmospheric Radiance

The final contribution to photon noise considered was radiance from the Earth's atmosphere. The atmosphere is composed a variety of mixed gases, most notably nitrogen, oxygen, argon, and water vapor. Behaving similarly to the ISM, gas molecules in the atmosphere are heated by the solar irradiance, as well as reflected radiance from the earth's surface, and then re-emit photons via various vibrational and rotational bands. These re-emitted photons compose the earth's atmospheric thermal radiance (ATR).

The atmospheric transmission and emission is generally extremely complex and vary widely with frequency and water vapor content. We generated atmospheric models using MODTRAN 5.2. Figure 2.4 depicts the BLING spectra for two ground sites which vary in altitude and atmospheric water content, as well as the BLING spectrum for airborne altitudes of 13.7 km (SOFIA) and 40 km (balloon). We modeled a 45 degree (relative to zenith) path through the atmosphere using the 1976 Standard Atmosphere model. We have used many other atmospheric models but do not include them as the changes are usually relatively minor, particularly given that we use the atmospheric PWV directly in the ground based models. As water is typically the critical issue in the far-IR, we use the best PWV quartile for the ground based sites. This will often imply that "clock integration times" will be substantially longer (often 4 times longer) compared to the best quartile we use. This depends on the band and weather however.

The two ground systems we modeled were Cerro Chajnantor in Chile and Dome A in Antarctica. Despite Dome A having a water vapor level seven times less than Cerro Chajnantor, the emission spectra were quite similar, though the transmission were very different in the far IR bands.

The PWV distribution for ground sites is highly variable and very seasonally dependent. Averages are simply that and any given year may be far worse or better than average. The weather at balloon altitudes is generally not an issue. It does exist but is not dominant as it is on the ground. On the ground, the PWV is sensitive to climate change and many ground based systems are expected to become worse with time.

Although the MODTRAN model can account for direct and scattered solar radiance and backscattering from the Earth's surface, this dramatically increases the specificity of the model (i.e. a need for time, location and day of year of the observation). Instead, we assumed the measurements were taken at night, as would be done where maximum sensitivity is necessary, and thus used the thermal radiance model. Nonetheless, the effect of solar scattered radiance and Earth backscatter is relatively small in the far-IR and only becomes significant in the near to mid IR and of course the visible. The latter will be considered in a future paper.

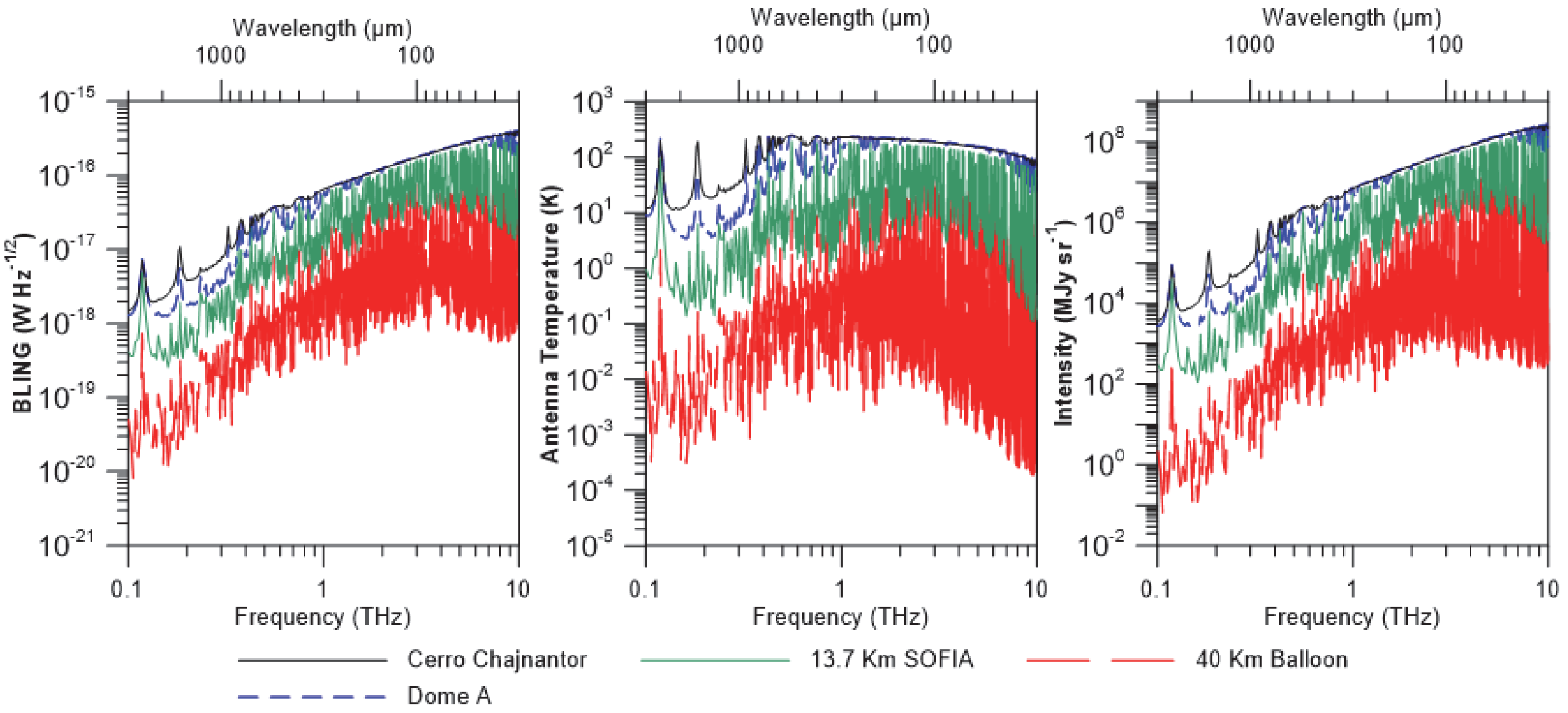

**Figure 2.4**: Atmospheric BLING, antenna temperature and intensity spectra for four atmospheric profiles. Despite Dome A having a sevenfold improvement in PWV over Cerro Chajnantor, there is minimal difference between them in their respective BLING spectra. However, the transmission is very dependent on the PWV and thus very different between the sites. The 40 km balloon model yields a BLING spectrum that varies widely with wavelength as the atmospheric lines are resolved due to the extremely low PWV, and minimal pressure broadening and is orders of magnitude lower (particularly at the deep minima) than that of the best ground sites.

## 2.4 Cumulative BLING

The total BLING for a telescope system is calculated by adding the constituent BLINGs in quadrature. In the case of optically thick components, such as the Earth's atmosphere, we attenuate the extraterrestrial BLING (and signal) as appropriate before calculating the total.

$$\zeta_{tot} = \left(\sum_i {\zeta_i}^2\right)^{\frac{1}{2}} \qquad \text{(eqn 2.18)}$$

In general, the integration time is proportional to the square of the BLING and hence minimizing the BLING is critical to maximizing the mapping speed. Figure 2.5 depicts cumulative BLING spectra from all sources for each telescope model. Where applicable, extraterrestrial noise sources were attenuated by the atmospheric transmission model (see section 3.2), although this effect is generally minimal. As expected, for a cold space mission, the extraterrestrial backgrounds dominate. For ground observations, the noise is dominated by the emissions of the warm atmosphere. In the case of high-altitude balloons, the CMB is the most significant source of noise for frequencies less than 0.3THz, whereas atmospheric lines, in combination with the warm mirror dominate at higher frequencies. In order to become limited in the far-IR by atmospheric and extraterrestrial sources the optics will typically have to be cooled to less than ~30 K, although this varies with wavelength. As integration time to a given SNR is typically proportional to $\zeta_{tot}^2$ it is critical to keep $\zeta_{tot}$ as low as possible. A factor of 20 in $\zeta_{tot}$ can turn a day into a year of observing. We will see the effect later when we compare sites and systems.

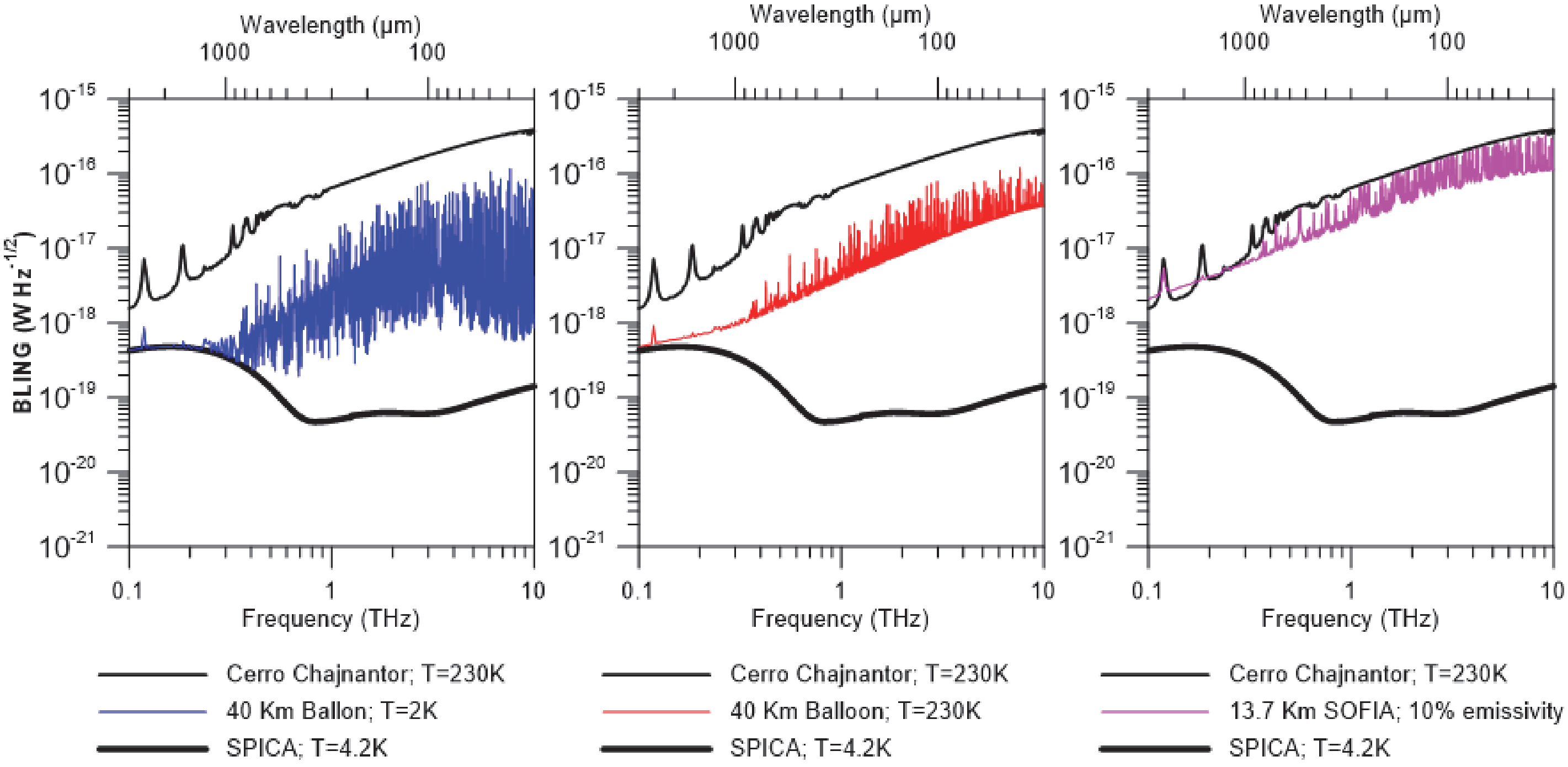

**Figure 2.5:** Cumulative BLING for four potential systems. The space based model (SPICA) is the lowest due to its lack of atmospheric noise. The Cerro Chajnantor and SPICA curves are shown in all panels to act as reference.

# 3. Signal

## 3.1 SED of Target Galaxies

The expected signal emitted by the target galaxy is the most important quantity to understand when making astrophysical observations. The galactic emissions of interest are in the submillimeter and far-IR waveband. Coincidentally, the galaxies of interest also contribute to the CIB.

Blain et al. (2002) developed a method for modeling the SED of far-infrared galaxies. We have used this method, as well as measured SED's, to generate our own models. About 99% of the energy released by galaxies in the submillimeter and far-IR wavebands is produced by dust grain thermal emission; the remainder is accounted for by fine structure atomic and molecular rotational line emission. Dust emits a relatively featureless (in terms of sharp lines) modified blackbody spectrum. We do not consider polycyclic aromatic hydrocarbons (PAH) here. The minimum parameters necessary to describe this spectrum are a dust temperature $T_d$ and an emissivity function $\varepsilon_\nu$ (Blain et al. 2002). In a given galaxy there will be a distribution of dust temperatures, corresponding to the structure and environment of each grain. For our purposes, we took $T_d$ to be the temperature of the coolest grains, that significantly contribute to the submillimeter and far-IR energy output of the galaxy. We focus on the observation of distant galaxies, for which there are very few spatially and spectrally resolved images, so we assumed a volume averaged emissivity function of frequency. A future paper will consider line emission. We model,

$$\varepsilon_\nu \propto \nu^\beta \qquad \text{(eqn 3.1)}$$

The value of $\beta$ is usually assumed to be between 1 and 2. The spectral energy distribution (SED) of the dust emission can now be expressed as $f_\nu$.

$$f_\nu \propto \varepsilon_\nu B_\nu \qquad \text{(eqn 3.2)}$$

$B_\nu$ is the Planck function $\left[\frac{W}{m^2 Hz\ sr}\right]$,

$$B_\nu = \left(\frac{2h\nu^3}{c^2}\right)\left(\frac{1}{e^x - 1}\right), \qquad \text{(eqn 3.3)}$$

where $x$ is defined in eqn 2.3.

It is important to note that this model is valid only in the submillimeter to far-IR wavebands. In the mid-IR and near-IR wavebands, dust grains with a temperature greater than $T_d$ and stellar emission will prevent the SED from dropping with a Wien exponential (Blain et al. 2002). In these bands it is instead reasonable to model the SED as a power law,

$$f_\nu \propto \nu^{\alpha_{high}}. \qquad \text{(eqn 3.4)}$$

At longer wavelengths, the SED once again deviates from a modified blackbody function. The slope of the SED typically changes abruptly at about 3 mm, where the dominant contribution shifts from thermal dust emission to synchrotron radio emission (Blain 1999). Synchrotron radiation is generated by the acceleration of ultra-relativistic charged particles through magnetic fields, and obeys a power law function of frequency,

$$f_\nu \propto \nu^{\alpha_{radio}}, \qquad \text{(eqn 3.5)}$$

The SED model we used for all frequencies of concern can now be expressed as a piecewise function.

$$f(\nu) \propto \begin{cases} \nu^{\alpha_{radio}},\ \nu < \nu_{radio} \\ \nu^\beta \left(\frac{2h\nu^3}{c^2}\right)\left(\frac{1}{e^{\frac{h\nu}{k_B T_d}} - 1}\right),\ \nu_{radio} < \nu < \nu_{midIR} \\ \nu^{\alpha_{high}},\ \nu_{midIR} < \nu \end{cases} \qquad \text{(eqn 3.6)}$$

$\nu_{midIR}$ is the transition frequency from the modified blackbody to the high frequency inverse power law. $\nu_{radio}$ is the transition frequency from the modified blackbody to the low frequency power law . These are found by fitting observed data.

We considered two examples of luminous infrared dusty galaxies, NGC 958 and Markarian 231 (Mrk 231). NGC 958 is an example of a cooler, less luminous galaxy, and Mrk 231 a hotter, brighter galaxy. We fitted the model to the SED observations from the NASA/IPAC Extragalactic Database as summarized in Table 3.1 and shown in Figure 3.1.

| | NGC 958 | Mrk 231 |
|---|---|---|
| SED radio-frequency emissivity exponent $\alpha_{radio}$ | -0.200 | -0.365 |
| SED Dust Temperature $T_d$ | 26.9 K | 50.4 K |
| SED Planck-region emissivity exponent $\beta$ | 1.28 | 1.21 |
| SED IR emissivity exponent $\alpha_{high}$ | -2.04 | -1.56 |
| Measured redshift | 0.0196 | 0.04217 |
| Transition frequency $\nu_{radio}$ | 0.14 THz | 0.24THz |
| Transition frequency $\nu_{midIR}$ | 3.2 THz | 4.8THz |
| $R^2$ fit of radio frequency power law | 0.870 | 0.877 |
| $R^2$ fit of modified blackbody | 0.999 | 0.991 |
| $R^2$ fit of high frequency power law | 0.983 | 0.959 |

**Table 3.1:** Fitted SED template parameters.

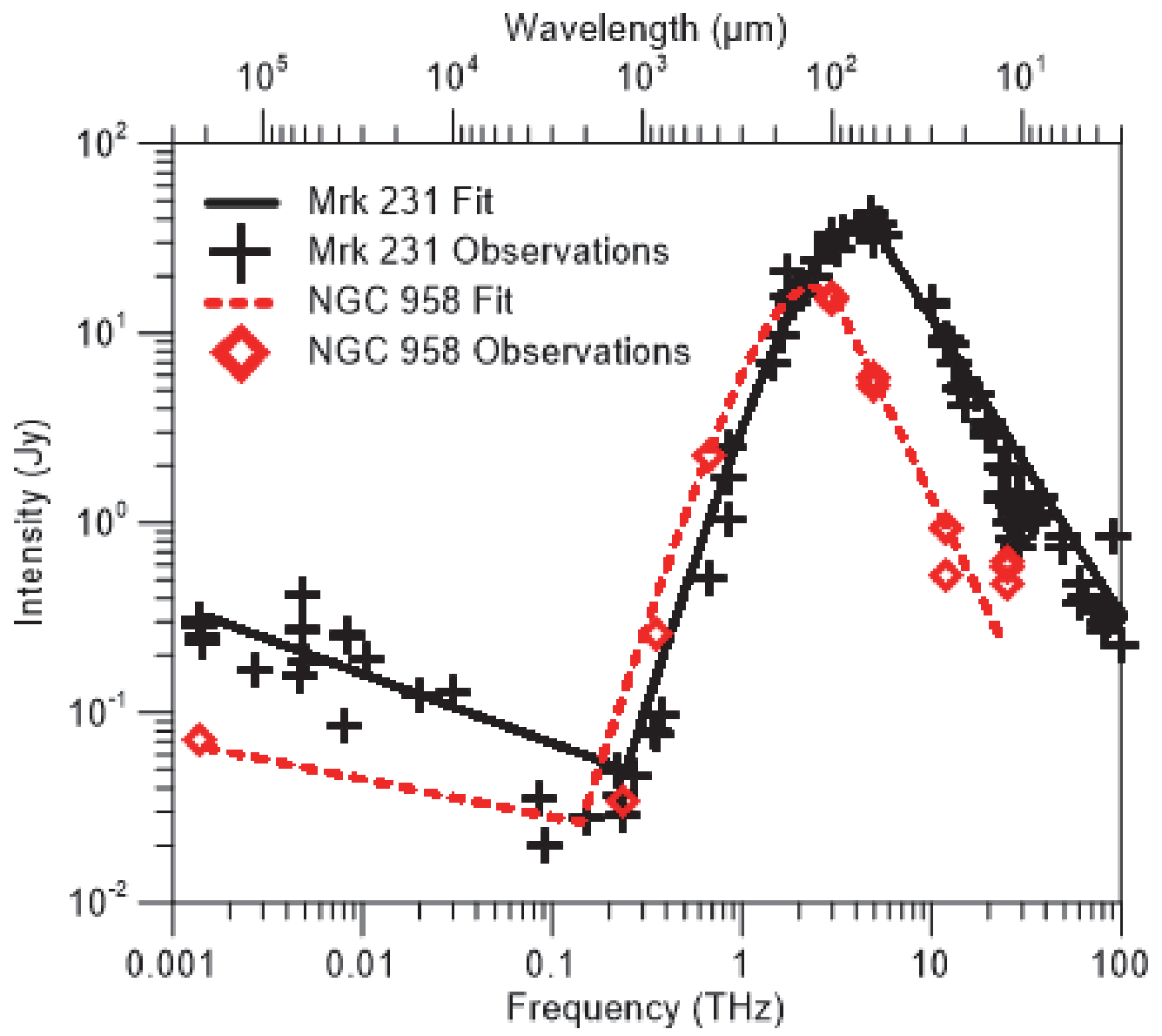

**Figure 3.1**: Fitted SED model and observed data from the NASA/IPAC Extragalactic Database of NGC 958 and Mrk 0231 at their observed redshift.

NGC 958 has a measured redshift of 0.0196 while Mrk 0231 has a measured redshift of 0.0422. In order to compare and model galaxies at different redshifts, the SED must be scaled. The luminosity distance $D_L$ of an object is defined by the relationship between the observed bolometric (i.e. integrated over all frequencies) flux $S$ $[W/m^2]$ and intrinsic bolometric luminosity $[watts]$.

$$S = \frac{4\pi D_L{}^2}{L} \quad \text{or} \quad D_L = \sqrt{\frac{L}{4\pi S}} \qquad \text{(eqn 3.7)}$$

The source function of a galaxy varies widely with frequency, rendering a bolometric approach less useful , and thus differential forms of $S$ and $L$ are used, $S_\nu$ $[W/m^2Hz]$ and $L_\nu$ $[W/Hz]$. As a result, the k-correction must be applied to either the flux or luminosity because the redshifted object is emitting flux in a different band (in its rest frame) than we are observing. We can use various cosmological models but for simplicity we will describe results for the benchmark model shown in Table 2.1.

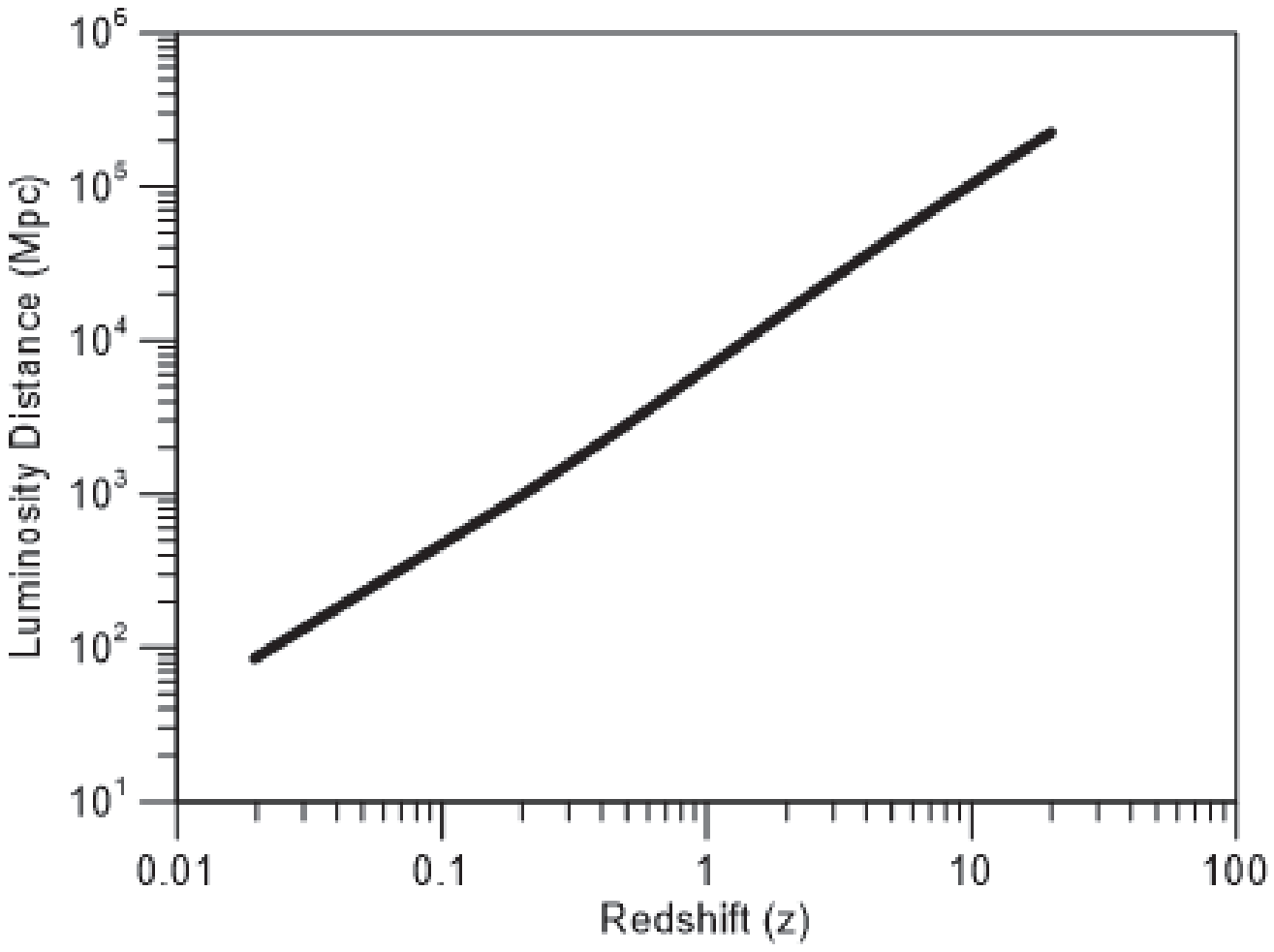

**Figure 3.2**: Luminosity distance as a function of redshift for the Benchmark Model of the Universe, used to generate our SEDs. Parameters assumed: $\Omega_M$ = 0.3, $\Omega_L$ = 0.7, $w_0$ = -1, $H_0$ = 70.

If a source at redshift $z$ emits at frequency $\nu_{emitted}$, we observe it at frequency $\nu_{obs}$.

$$\nu_{obs} = \frac{\nu_{emitted}}{1+z} \qquad \text{(eqn 3.8)}$$

For a bandwidth at the source $\Delta\nu_{\text{emitted}}$, between frequencies $\nu_{emitted,1}$ and $\nu_{emitted,2}$, the bandwidth at we observe is $\Delta\nu_{obs}$, between frequencies $\nu_{obs,1}$ and $\nu_{obs,2}$.

$$\Delta\nu_{\text{emitted}} = \nu_{emitted,2} - \nu_{emitted,1} = (1+z)\nu_{obs,2} - (1+z)\nu_{obs,1}$$

$$\Delta\nu_{obs} = \frac{\Delta\nu_{\text{emitted}}}{1+\text{z}} \qquad \text{(eqn 3.9)}$$

The observed differential flux at frequency $\nu$ when observing an object at redshift z, $S_{\nu,z}$ is related to the emitted differential luminosity $L_\nu$, by

$$S_{\nu,z} = \frac{(1+z)L_{\nu(1+z)}}{4\pi {D_L}^2} = \frac{(1+z)L_\nu emitted}{4\pi {D_L}^2} \qquad \text{(eqn 3.10)}$$

We start with the observed flux $S_{\nu_0,z_0}$ and then compute $L_\nu emitted$ where $\nu_{emitted} = (1+z_0)\nu_0$. Here $\nu_0 = \nu_{obs}$, and $z_0 = z_{obs}$. We thus calculate

$$L_\nu emitted = L_{(1+z_0)\nu_0} = \frac{4\pi {D_L}^2 S_{\nu_0,z_0}}{1+z_0}$$

where $D_L$ is the luminosity distance at the observed redshift. We then compute $S_{\nu,z}$ at any given redshift from equation 3.10 (Hogg 1999). The factor of (1+z) is due to the compression of bandwidth due to redshifting.

Another way to visualize this is to consider a given rest frame emission frequency and redshift this to different redshifts $z_1$ and $z_2$ and observed frequencies $\nu_1$ and $\nu_2$ . To find the spectrum at some $z_2$, we must use equation 3.10 twice and take the ratio of the results.

$$S_{\nu 1,z_1} = \frac{(1+z_1)L_\nu emitted}{4\pi {D_{L_1}}^2} \qquad S_{\nu 2,z_2} = \frac{(1+z_2)L_\nu emitted}{4\pi {D_{L_2}}^2}$$

$$\frac{S_{\nu 2,z_2}}{S_{\nu 1,z_1}} = \left(\frac{1+z_2}{1+z_1}\right)\left(\frac{D_{L_1}}{D_{L_2}}\right)^2 \qquad \text{(eqn 3.11)}$$

since $L_{\nu_1(1+z_1)} = L_{\nu_2(1+z_2)=} L_\nu emitted$, as we assumed $(1+z_1)\nu_1 = (1+z_2)\nu_2 = \nu_{emitted}$.

If we are interested in observing the galaxy at some redshift $z_2$, and we know the functional form of $S_{\nu_1}$ from equation 3.6 we can map $S_{\nu_1}$ into the $\nu_2$ frame:

$$\nu_1 = \left(\frac{1+z_2}{1+z_1}\right)\nu_2$$

$$S_{\nu 2_{,z2}} = \left(\frac{1+z_2}{1+z_1}\right)\left(\frac{D_{L_1}}{D_{L_2}}\right)^2 S_{\nu 1_{,z1}} \quad \text{or} \quad S_{\nu,z2} = \left(\frac{1+z_2}{1+z_1}\right)\left(\frac{D_{L_1}}{D_{L_2}}\right)^2 S_{\left(\frac{1+z_2}{1+z_1}\right)\nu_{2,z1}} \qquad \text{(eqn 3.12)}$$

Equation 3.12 incorporates the effects of the "K correction" which can also be an inverse K correction depending on whether we are redshifting from a "falling with frequency spectrum" (normal K correction) or a "rising with frequency spectrum" (inverse K correction). Because the absolute change in shifted frequency is proportional to the emitted frequency, the bandwidth of the observed signal is also redshifted and thus narrowed. Because of this compression, the differential flux, $S_{\nu_2}$ gains an extra factor of $\left(\frac{1+z_2}{1+z_1}\right)$. Additionally, since $S_{\nu_1}$ is mapped into the $\nu_2$ frame, if $\nu_2$ is on the Rayleigh-Jeans portion of the SED (rising spectrum) with emissivity $\sim\nu^\beta$ , then we will get an inverse K correction and the source will appear to be brighter than naively expected from distance alone. In this case $S_{\nu_2}$ will scale as $\sim\left(\frac{1+z_2}{1+z_1}\right)^{3+\beta} * \left(\frac{D_{L_1}}{D_{L_2}}\right)^2$ and will increase significantly for large z. Two powers come from the quadratic Rayleigh Jeans spectral increase and one power comes from redshifted bandwidth compression with the addition power of $\beta$ coming from the dust emissivity. As mentioned, $\beta$ is typically between 1 and 2. The inverse K correction provides an extremely powerful probe of high redshift dusty galaxies that would otherwise be much fainter without this effect. A similar effect happens with the redshift on any rising spectrum SED. For example, this often occurs in the near IR for the short wavelength "stellar" SED component. The inverse K-correction NGC 958 template using the methods outlined above is shown in Figure 3.3 and the correction applied to the template SEDs are shown in Figure 3.4.The inverse-K correction greatly benefits the observation of millimeter-wave and microwave galaxies with the peak effect near 100 GHz for $z{\sim}10$ galaxies with typical dust temperatures. The overall effect is that high redshift galaxies in the far-IR appear much brighter than naively expected from a pure inverse square law analysis. The redshifting of the dust peak to lower frequencies can result in a higher apparent intensity at $z = 2$ than $z = 1$ as the peak approaches the frequency being observed. This offers the possibility of a unique probe of very high redshift structures which can be brighter at $z{\sim}10 - 20$ than at $z{\sim}1$. We do not currently have a good understanding of the distribution of high redshift dusty galaxies, and thus this area is of great importance to explore.

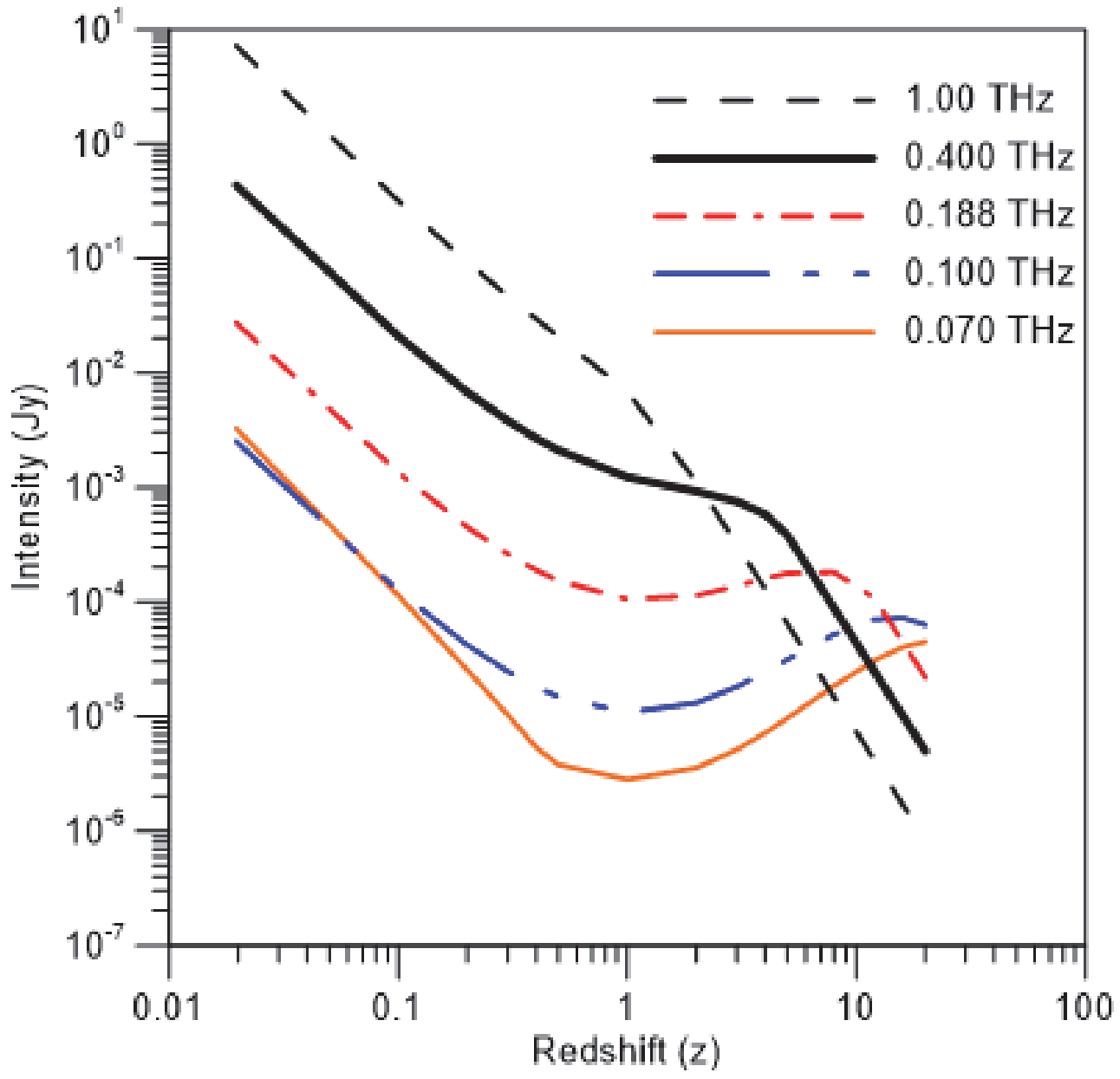

**Figure 3.3**: Intensity vs. redshift for the NGC 958 template. This highlights the effects of the inverse K-correction for frequencies 0.07-1THz for redshifts out to z=20.

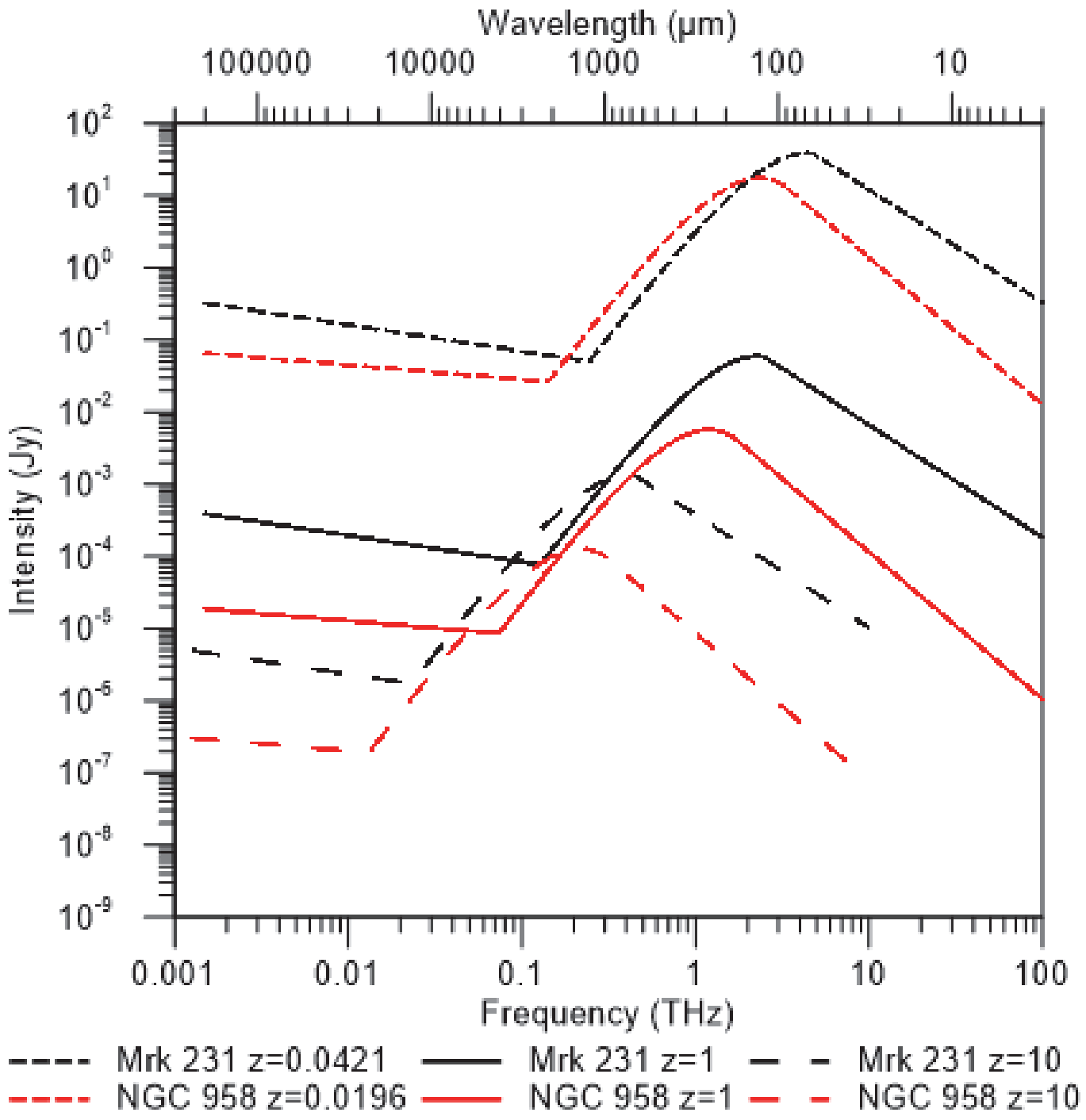

**Figure 3.4**: Redshifted template SEDs . Note the increased intensity at higher z for some cases.

## 3.2 Impact of Water Content on Atmospheric Transmission

In the wavelength range of 30-3000µm, water absorbs a significant amount of incident electromagnetic radiation. Photon energies in this wavelength range have energies that closely match the difference in energy levels for a myriad of vibrational modes of water vapor molecules (Chaplin 2011). Thus, when an incident photon encounters water molecule, there is a high probability of absorption. We have modeled all major atmospheric gasses with water vapor absorption typically being the dominant effect in the THz regime.

Due to the profound impact water content plays upon transmission in the far-IR and submillimeter band, it is important to minimize the precipitable water vapor (PWV) when searching for a new ground based site. The PWV represents the total amount of water in a column of atmosphere that could precipitate as rain. Since the water vapor absorption is extremely sharp in many parts of the spectrum, factors of 2 increase in PWV can lead to an order of magnitude or more increase in absorption. Thus it is critical to minimize the total PWV as much as possible. The search for the best ground site on earth has resulted in the discovery of a number of viable high and dry options a few of which are listed in Table 3.2. Domes A and C in Antarctica have extraordinarily low PWVs (Yang et al. 2010), but are extremely remote. The South Pole is also a potential option (Yang et al. 2010) but not nearly as good. Cerro Chajnantor is also notable as one of the best mid latitude sites with a best quartile PWV of 0.0732 g/cm$^2$ (Herter and Radford 2005). The CCAT project is an ambitious 25 m diameter telescope slated for Cerro Chajnantor (Gold 2007), being very good in the long wavelength far-IR transmission windows.

| % Time with ≤ PWV | Dome A (4100 m altitude) | Dome C (3250 m) | South Pole (2850 m) | CCAT proposed (5600 m) | Cerro Chajnantor (5100 m) | White Mountain (3800, 4000 m) | Mauna Kea (4100 m) |
|---|---|---|---|---|---|---|---|
| 25% | 0.010 | 0.015 | 0.023 | 0.36 | 0.073 | 0.115 | 0.10 |
| 50% | 0.014 | 0.024 | 0.032 | 0.67 | 0.098 | 0.175 | 0.15 |

**Table 3.2**: Precipitable Water Vapor contents used for various ground sites $[g/cm^2]$. Dome A, Dome C, South Pole, and Mauna Kea values are from Yang et al. 2010. To convert PWV to mm, multiply the value in $[g/cm^2]$ by 10. Values for Cerro Chajnantor are from Herter and Radford 2005. We note that the siting for the CCAT telescope will likely be at 5600m with some evidence for the best quartile PWV being closer to 0.036 $g/cm^2$ (Radford Dec 2012). In the text when we refer to Cerro Chajnantor we use the PWV quartiles from Herter and Radford 2005 which are close to the ALMA site PWV values. White Mountain values are from Marvil et al. 2005. As with all mid-latitude sites including Cerro Chajnantor, Mauna Kea, and White Mountain, the quartiles can vary significantly from day to day and from year to year, making scheduling of the deepest surveys more complicated.

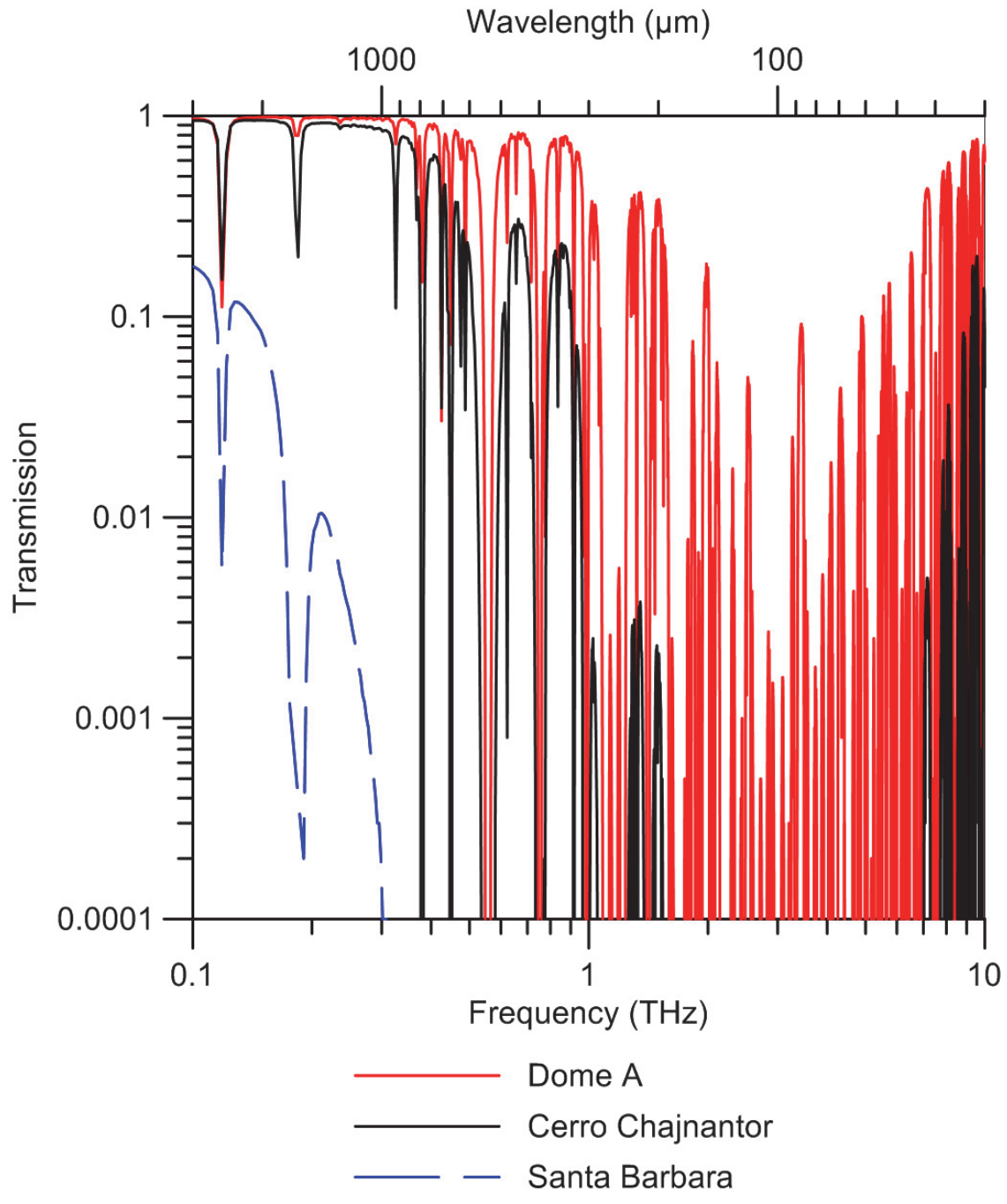

**Figure 3.5**: Transmission spectra for three ground sites, for the best 25% of time, and an observation 45 degrees from Zenith. The 1976 Standard Atmosphere distribution was used for scaling PWV along the line of sight. Santa Barbara, California is included as a typical near-sea-level location. The Santa Barbara (sea level) PWV was chosen to be 3.0 $g/cm^2$. Note the general lack of transmission above 400 GHz, except in select windows. From 1-10 THz, ground based observations are extremely difficult except in the most extreme sites, and even then only in relatively narrow windows.

It is apparent from Figure 3.5 that despite extremely low PWVs from the best ground based sites, transmission from these sites still drops significantly in the 1-10 THz range, with many frequency regions being nearly completely opaque. To maximize transmission, one must minimize the effects of water vapor. This can be achieved at very high altitudes from air borne systems and, of course, in space.

For airborne systems we model SOFIA as operating at 13.7 km and high-altitude balloons at 40 km. We only show the 1976 model atmosphere data here. The airborne transmission spectra are shown in Figure 3.6. Since the balloon flies 20+ km above the airplane, it is dramatically superior in transmission quality.

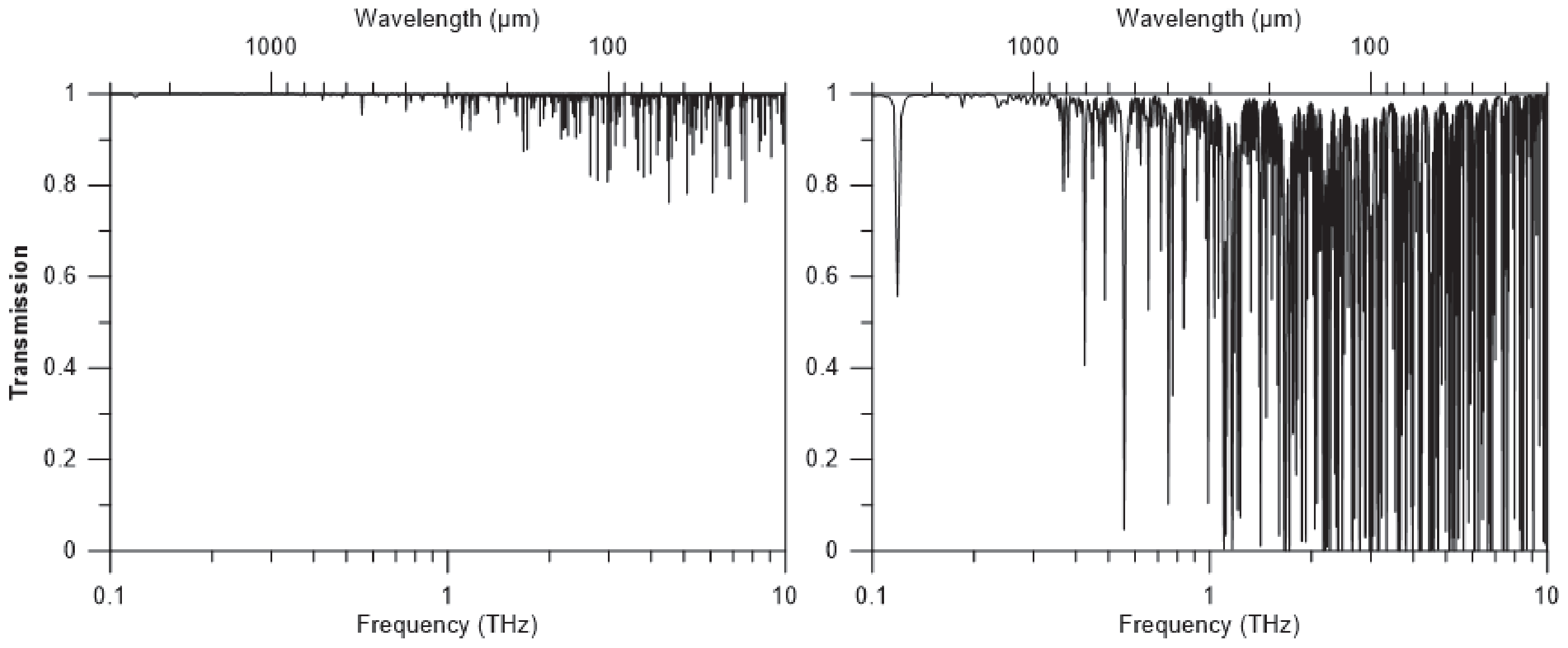

**Figure 3.6**: Transmission spectra for a 40 km altitude balloon (left) and 13.7 km aircraft (right), looking at 45 degrees from zenith using the 1976 Standard Atmosphere model. While the general trend of the 13.7 km transmission spectrum is vastly superior to the ground, there are still significant amounts of opaque regions which can be reduced or eliminated by going to higher altitudes.

## 3.3 Final Signal Power

The signal power a detector receives is dependent on the source intensity function $S(\nu)$, the transmission function $\tau(\nu)$, the diameter of the mirror $d$, and the bandwidth $\Delta\nu$.

$$P(\nu_0) = \pi\left(\frac{d}{2}\right)^2 \int_{\nu_0-\frac{\Delta\nu}{2}}^{\nu_0+\frac{\Delta\nu}{2}} \tau(\nu)S(\nu)d\nu \qquad \text{(eqn 3.13)}$$

In our case $S(\nu)$ is the flux in equation 3.12 and is in units of $[W/(m\hat{}2\ \ Hz)]$.

Signal power $[watts]$ spectra for our galaxy templates are presented in Figure 3.7 for several systems of varying aperture. Figure 3.7 shows that ground sites (which have the largest apertures) have the highest signal strength for frequencies less than 1 THz, though there are a few regions of heightened opacity. However, for 1-10 THz the significant opacity of the atmosphere causes the signal of the ground sites to fall well below that of the balloons and space observatory.

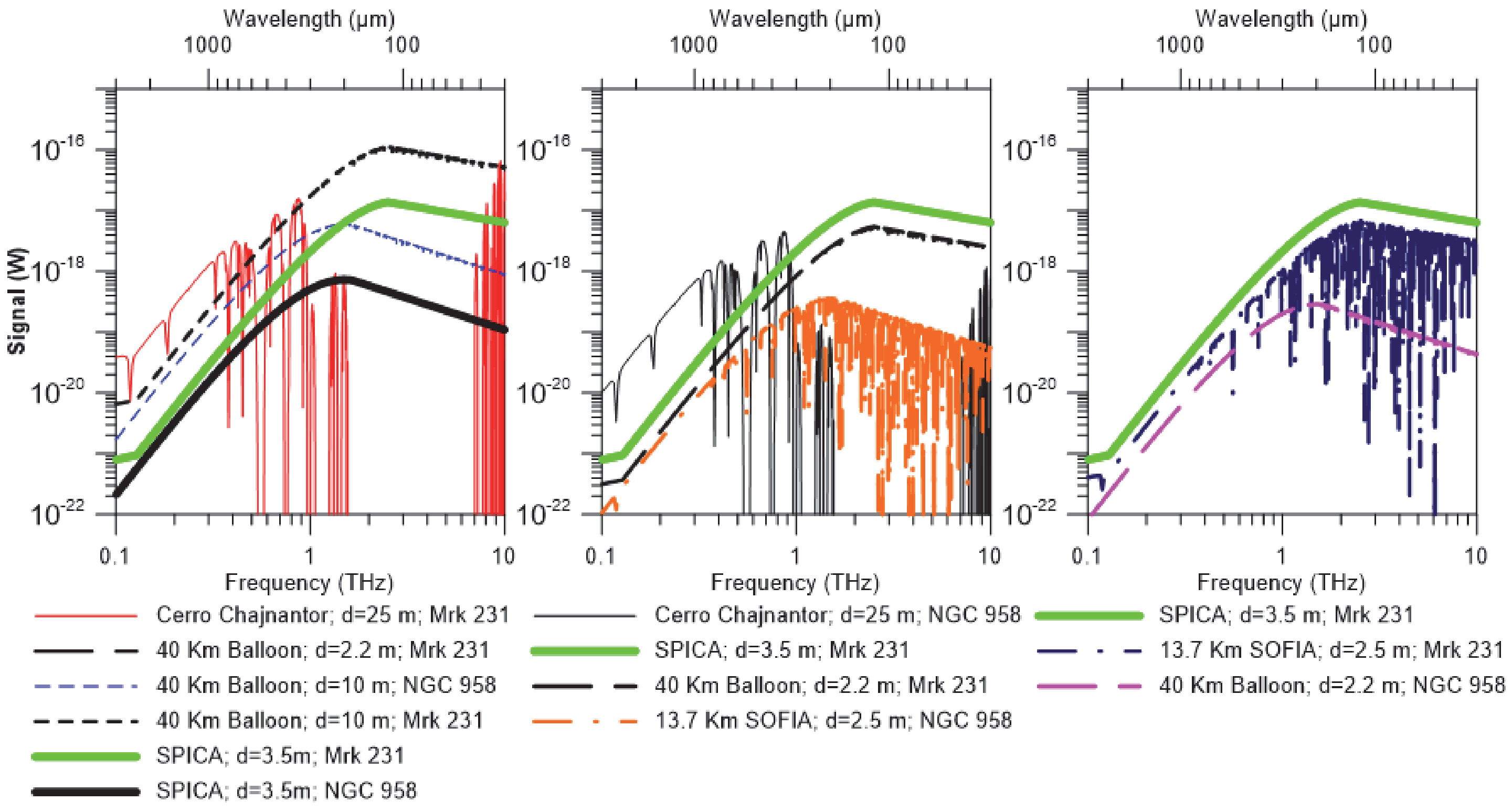

**Figure 3.7**: Signal spectra for systems observing both the Mrk 231 and the NGC 958 SED template when placed at z=1.

## 4. Integration Time

Signal and noise alone do not determine the performance of the system, but rather the ratio of signal to noise (SNR) is the primary measure of signal performance. SNR is a function of the time over which the signal and noise are integrated.

Integration time is the time it takes to achieve a particular signal to noise power ratio. In our case, we have noise expressed as BLING, which has units of $W/\sqrt{Hz}$. We first need to express noise as a power $P_N$,. Then we can solve for integration time t in terms of total BLING $\zeta_{tot}$, signal power $P_S$, and the desired signal to noise ratio $(SNR) = \sigma$.

$$\zeta_{tot} = P_N\sqrt{t} \qquad \Rightarrow \qquad P_N = \frac{\zeta_{tot}}{\sqrt{t}}$$

$$\sigma = \frac{P_S}{P_N} = \frac{P_S\sqrt{t}}{\zeta_{tot}} \qquad \Rightarrow \qquad t = \left(\frac{\zeta_{tot}\sigma}{P_S}\right)^2 \qquad \text{(eqn 4.1)}$$

Note the scaling dependence of integration time t goes as the square of the total BLING $\zeta_{tot}$, the square of the SNR, and inversely as the square of the signal. Since the observed signal is dependent on the overall transmission of the atmosphere, this will become critical when we calculate the required integration time for ground and airborne platforms to observe a given SNR.

The integration times for several scenarios is presented in Figure 4.1. Ground sites dominate despite atmospheric effects for frequencies less than 0.3 THz due to the large aperture size possible (25 m diameter assumed here). Once above the 0.3 THz threshold, the minimized noise and maximized signal of space based telescopes result in their clear dominance. However, the efficiencies of a 230 K, 10 m diameter telescope or a 2 K, 2.2 m diameter telescope flying on a 40 km balloon are also notable. These balloon scenarios present a viable alternative to space missions in some cases. The costs for balloon payloads are typically several orders of magnitude lower, are much faster to flight and can use the current state of the art in detectors as there is little technological "lock in time". Finally, we note that the SOFIA airborne telescope has high integration times, as a result of both being subject to atmospheric attenuation and noise, as well as being limited in mirror diameter. When the "as built" additional emissivity of SOFIA is added to the model, integration times rise by about one order of magnitude from the idealized case of emissivity from bulk metal only (Figure 4.4b). We note, however, that the SOFIA telescope is the only instrument already constructed and flying.

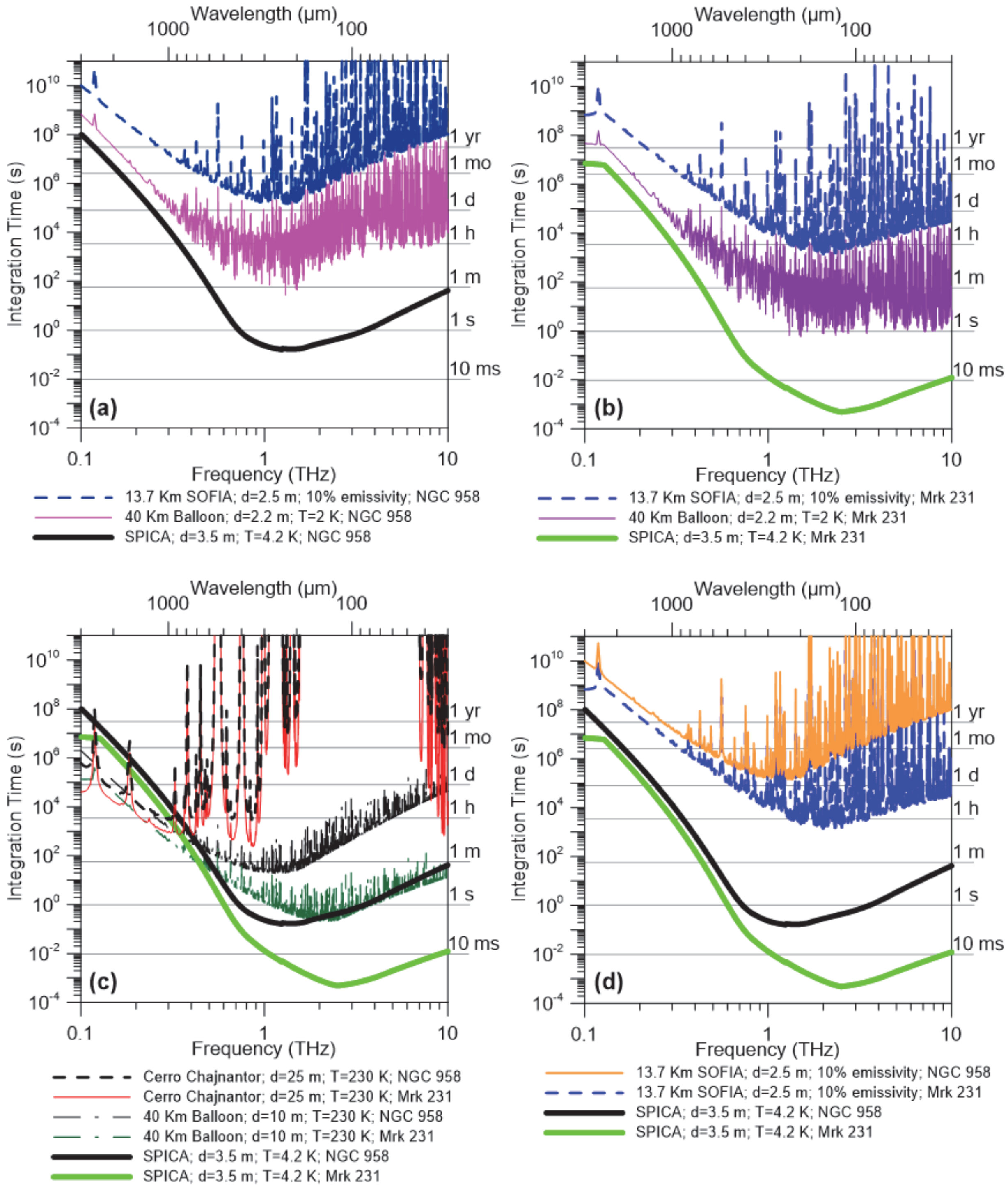

**Figure 4.1**: Integration time versus frequency for a four possible telescope systems observing a z = 1 galaxy with two SED templates and R= 1000. Graph (a) compares high-altitude, small-aperture platforms observing the NGC 958 template. Graph (b) depicts the same observing the Mrk 231 template. Graph (c) depicts the ground telescope and the large-aperture balloon models. Plot (d) compares the two SED templates on the SOFIA model. Ground sites use the best PWV quartile and thus "clock integration time"

is typically four times larger than shown. We have used a rather optimistic far-IR emissivity of 10% for SOFIA.

Analysis of Figure 4.1 shows distinct regions where different systems excel. When comparing platforms it is important to consider the logistics of each. For example, it is reasonable to run a ground site for many days of a year, but it is not reasonable to do so with a balloon or an aircraft. A short duration balloon mission (~12-24 hrs) may run only a few days per year, and so we would have to make a compelling argument in favor of a balloon in this case; it should have an integration time of at least 100 times less than the ground site. In contrast, if a balloon has an integration time three orders of magnitude less than a ground telescope, it means the balloon can return the same data in one day as nearly three years of ground site operation and thus is highly favorable, although confusion limits are often a serious issue and aperture size is needed in these cases. The total amount of mass that can be lofted for detectors is also a serious issue and balloon missions need special low mass instruments. There are clear tradeoffs. Figures 4.1a-d make this comparison. Long duration balloon flights of about a month are now possible and 100 day missions look feasible in the future. This could drastically change the science return in these cases.

At frequencies below 0.4 THz, the large diameter possible with ground based telescopes dominates as a space-based mission, such as a SPICA class mission, would be up to 25-100 times slower in mapping and have much less angular resolution than a ground telescope, and would generally be far costlier (Fig 4.1c). If spaceflight is not a feasible option, then the large aperture, uncooled balloon-borne telescope is the good choice from 0.4 THz to 6 THz, delivering excellent mapping speeds. At 100 µm (3 THz), near the peak of emission for a nearby dusty galaxy, for example, a 25 m ground based telescope at a mid latitude sites could require years of observations to match a few minutes from a balloon measurement. From 6 to 10 THz the cooling of the mirror becomes beneficial for balloon borne measurements, though gains are modest for many parts of the spectrum, except in the deep troughs of the atmospheric transmission lines where it can be very important. Here, the cooling of the mirror allows the atmosphere to dominate. Cooling of a mirror is space is possible (Herschel, JWST, SPICA etc) but on a balloon it is quite difficult, though smaller mirrors can, and have been, cooled. Cooling a 2.2 m, like the UCSB systems, has not been done for balloon flights so far.

When comparing the two SEDs, the higher luminosity of the Mrk 231 template results in a lower integration time. Relative to the NGC 958 template, a vast increase in luminosity occurs after the NGC 958 dust peak at about 1.5 THz (z=1, 27 K dust) through the Mrk 231 peak at about 2.5 THz (z=1, 50 K dust). This results in integration times around 10 times lower for Mrk 231 at about 0.8 THz, reaching a peak of 3 orders of magnitude around 2-10 THz. This range is extremely difficult for ground based telescopes. For SOFIA, observing an Mrk 231-like object at z=1 near its peak emission requires an integration time of less than an hour, whereas an observation at the same frequency of an NGC 958-like object at z=1 requires days. This shows the tremendous sensitivity to the host galaxy SED, and thus makes blind surveys much more complicated to predict without more knowledge of the host galaxy distribution. It is important to note that the actual PWV distribution (not just the average quartiles) is also very important as the best 5% or 10% PWV can be extremely important in searching for the faintest objects.

Note that for ground based sites we show the best quartile PWV, so the long-term integration time ("clock time") for deep surveys is typically 4 times larger than shown. For example, in the best region of the 350 µm (ground based) window, one day on a balloon with a warm 10 m telescope is equivalent to more than a year observing on the ground with a 25 m telescope at an excellent mid latitude site. Outside the limited ground windows, the comparison is far more severe. For example, in the 100-200 µm band, a balloon-borne 10 m warm telescope is roughly a million times faster than a 25m ground telescope at a mid latitude site. Similarly a day on a balloon with a 10 m warm telescope is equivalent in sensitivity to a decade of dedicated SOFIA flights (assuming 100 SOFIA flights per year at 10 hours per flight). This also assumes an idealized SOFIA telescope emissivity. For the current SOFIA telescope the number is far worse. Even a cooled 2.2 meter balloon telescope does extremely well in the far-IR, as is shown in Figure 4.1. Long duration ballooning (LDB) and ultra-long duration balloon missions (ULDB) in the far IR become quite compelling.

Another critical issue to be considered is that the PWV slant path increases rapidly with the observation angle from zenith, and since the integration time is scaling as the square of the exponential drop in transmission due to PWV, the fraction of the sky available for deep surveys is much broader for systems with extremely low PWV such as balloon borne systems. A site such as the South Pole has in the consideration that targets are at the same elevation angle over time, but sky coverage is more limited as well, due to rapidly increasing slant PWV as you move off zenith. Balloon borne systems on the other hand cannot look "up" as the balloon is above.

In Figure 4.2 we show the comparison between Dome A and Chajnantor for the measured PWV in the best quartile in Table 3.2. Note how the significantly lower PWV at Dome A makes tremendous difference in the sub mm bands. The actual PWV being so highly weather and season dependent makes site comparisons much more complex and highly debated but general trends can be seen. One important issue is that mid latitude sites tend to be more dependent on "wet" and "dry" years. In any siting decision practical considerations must often override “best optical site” considerations as infrastructure is crucial to a successful program.

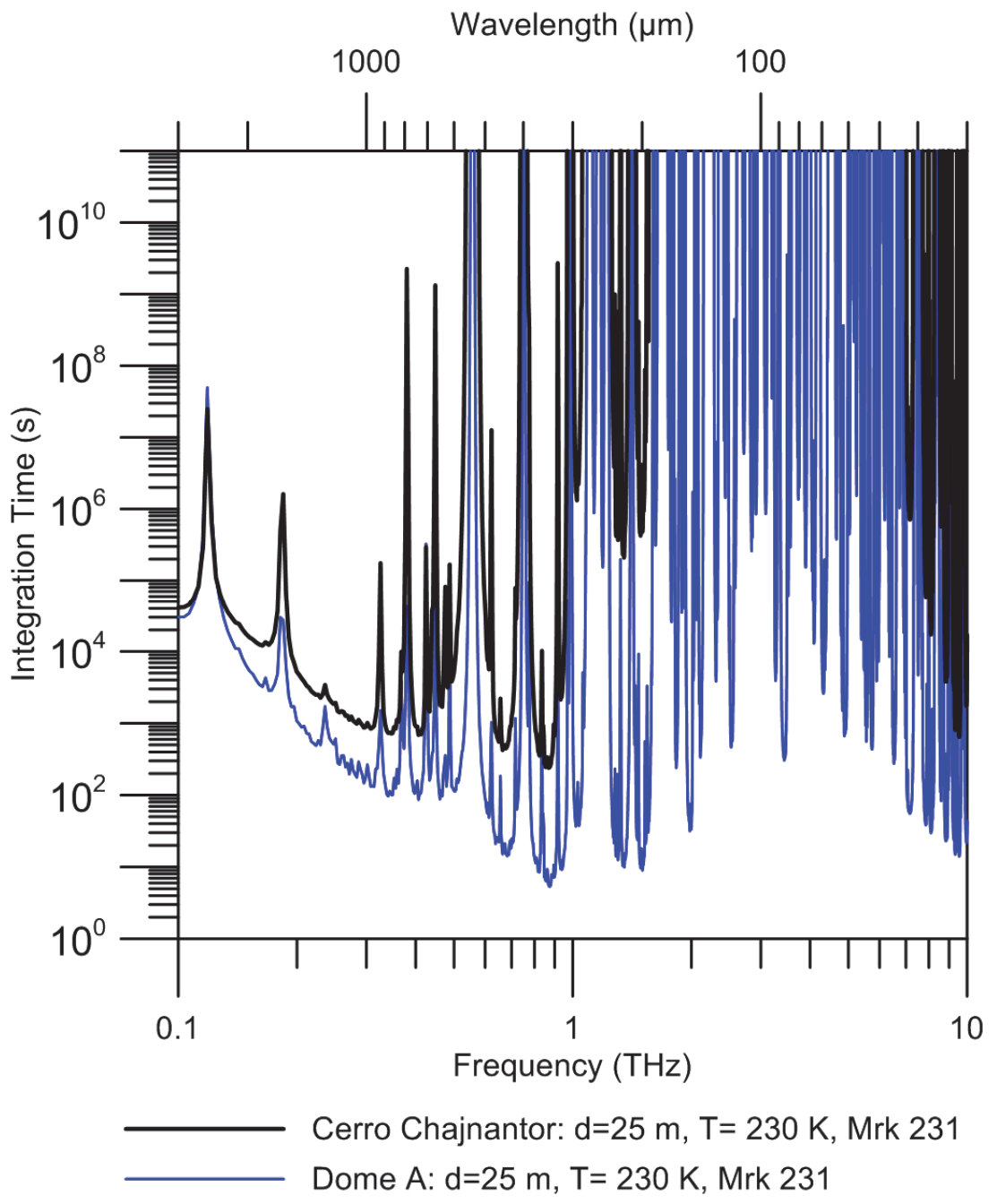

**Figure 4.2**: Comparison between two excellent ground based sites at Dome A in Antarctica and Cerro Chajnantor in Chile with 25 meter telescopes viewing NGC 958 template at z=1 for SNR=5 in two polarizations for the nominal 25% PWV. Notice the significant difference caused by the lower PWV at Dome A. Elevation angle for observation is 45 degrees. The proposed CCAT site at 5600 m will be between these two with estimated 25% PWV of 0.36mm, versus 0.10 mm for Dome A and 0.73 mm for Cerro Chajnantor.

In Figure 4.3 we compare our results with those of the ATM atmospheric modeling code (Pardo et al. 2001). For this comparison we model the PWV from 0.01 mm to 1.4 mm. The lowest values of PWV (<0.1 mm) are not realistic for Cerro Chajnantor. They become more relevant for Dome A though even here 0.01 mm is not within reach. The Modtran model shown is run for a 1976 atmosphere distribution though we also ran it for Tropical, Sub Arctic Winter and Summer and Mid Latitude Winter and Summer with similar results. The agreement is reasonable for the lower PWV values and diverges at higher values. Note the roughly exponential drop in transmission with PWV. Each window is in general unique and will normally have such an exponential behavior. Since integration time is roughly inversely proportional to the square of the signal and hence the square of the transmission, the PWV becomes a critical limiting factor in mapping speeds. This plot is about the same for each high altitude ground site as the transmission largely depends on PWV, not altitude. As we move to shorter wavelengths the exponent in the roughly exponential behavior with PWV becomes larger, and the effect of the site PWV becomes even more pronounced. We generally calculate for an assumed observation elevation angle of 45 degrees rather than at zenith as vertical is not realistic in most cases. It is not an insignificant difference at the shorter wavelengths where the exponential drop in transmission steepens, and hence a factor of roughly $\sqrt{2}$ in slant length PWV going from zenith to 45 degrees can have a substantial effect in integration time required. The tracking of a source for long integrations for non polar sites requires changing elevation, which further complicates matters.

The specific site chosen should be driven by the science requirements as well as practical issues. Less exotic sites can be perfectly adequate depending on the science target. A much more complex issue is the stability of the atmosphere. The fluctuations caused by very rapid variations in the temperature and water vapor (effectively rapidly moving "clouds") are often the limiting factor in observations in many systems attempting to observe extremely low flux levels, which require long integration times. The extreme case is observation of the CMB, for which some searches integrate for years (deep B mode searches for example) and site stability and transmission are critical. Typically the lower PWV sites also have better stability, particularly sites with very stable meteorology such as the Antarctic sites. This is a vigorously debated issue and one we will not discuss further, except to say that the lack of hard data often pushes observations toward the more extreme sites whether truly needed or not. The latter point is that barring proper data on fluctuations versus scale, frequency, and season, the best way to determine whether a site is suitable is often to do the measurement there. Since one does not usually have the ability to do the same measurement at a variety of sites and compare, it is difficult to choose a-priori which site is the most "cost effective" to get the science done. This is most clearly seen in some measurements done from the ground, balloon and space all targeting similar science with vigorous arguments as to the efficacy and "need" of each.

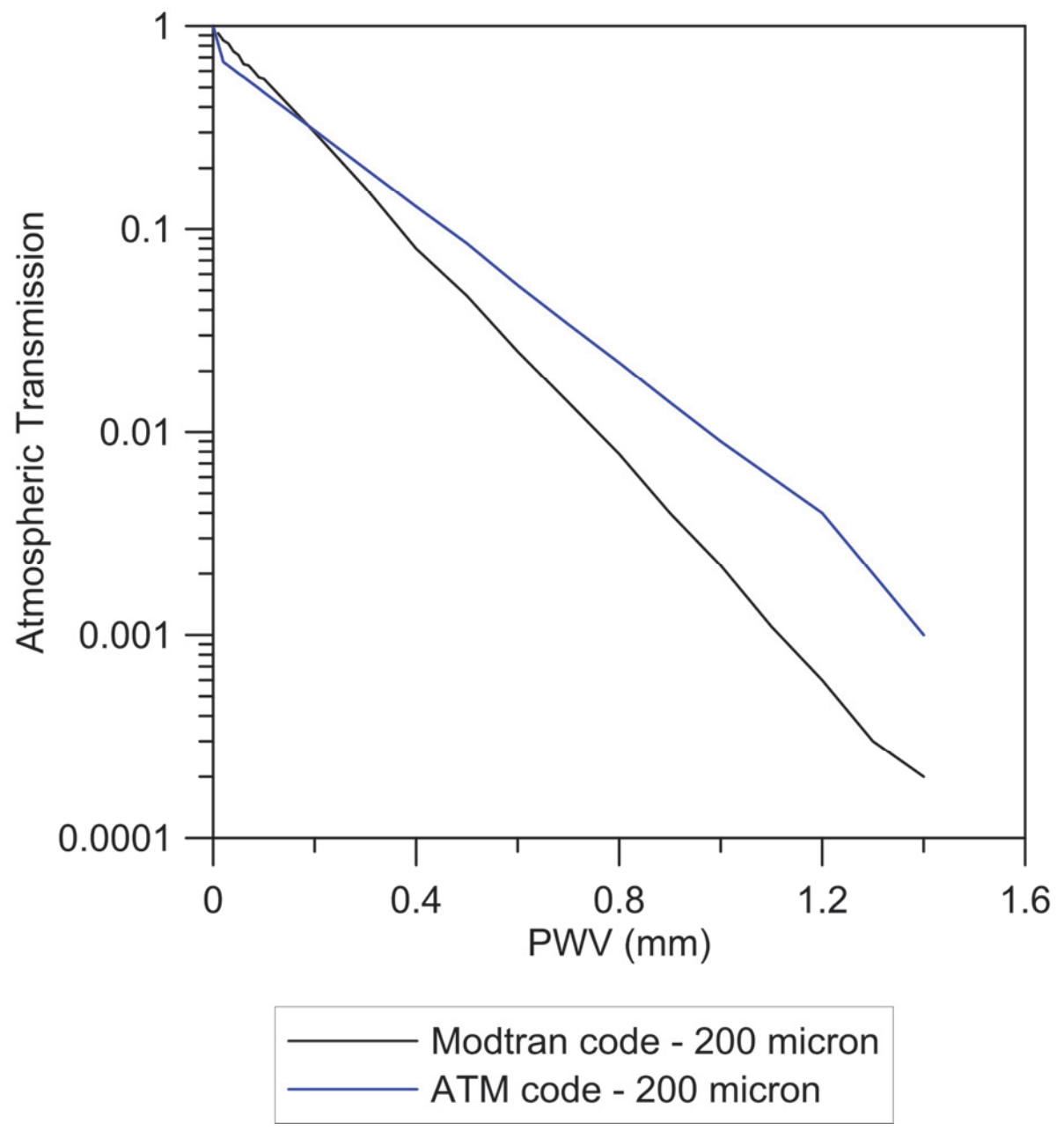

**Figure 4.3:** Comparison of ATM code (Pardo et al. 2001) versus our model calculations. The comparison is done at zenith in the 200 micron window, or 1.5 THz, (near max freq for ATM) and nominally for the Cerro Chajnantor APEX site (5.1 km altitude). In the critical PWV range from 0.1 to 0.5 mm agreement is reasonable given the complexity of the atmospheric radiation transfer codes involved.

It is important to note that missions are generally optimized for specific regions. SOFIA was primarily intended for near and mid IR observations and operates extremely well in that region. Use in the far-IR is not optimum, nor was it intended to be. In Fig 4.4b we compare the as built SOFIA with 10% emissivity and an idealized equivalent polished aluminum bulk emissivity model showing a significant reduction in integration time possible in such a case. In Fig 4.4a we compare a balloon borne 2.2 meter (like the UCSB Carbon Fiber systems we have flown for 15 years) running at 230 K (normal "warm" balloon temperature) and a cryogenic version running at 2 K. Note the greatly reduced integration times for the cryogenic case at the shorter wavelengths due to the reduction in optics emission. The balloon case is dramatically better over much of the far IR, especially in the case of a cryogenic balloon borne mirror. We note reduction in mirror temperature to 2 K is not needed in general as below 30K is usually sufficient (see Fig 2.2 and 2.3). Cooling a 2.2 m balloon mirror is not a trivial task, however, due to atmospheric condensation. While cryogenic open aperture instruments have been flown on some balloon payloads (e.g. ARCADE II) it is complex to do so and no cryogenic meter class mirror (let alone 2 meter class) has been flown yet. A segmented 10 meter balloon borne telescope for far IR work looks feasible based on the UCSB work on CFRP telescopes and has been proposed by the UCSB group as the TMMT program, but would clearly be a very challenging project. The rewards would be remarkable however.

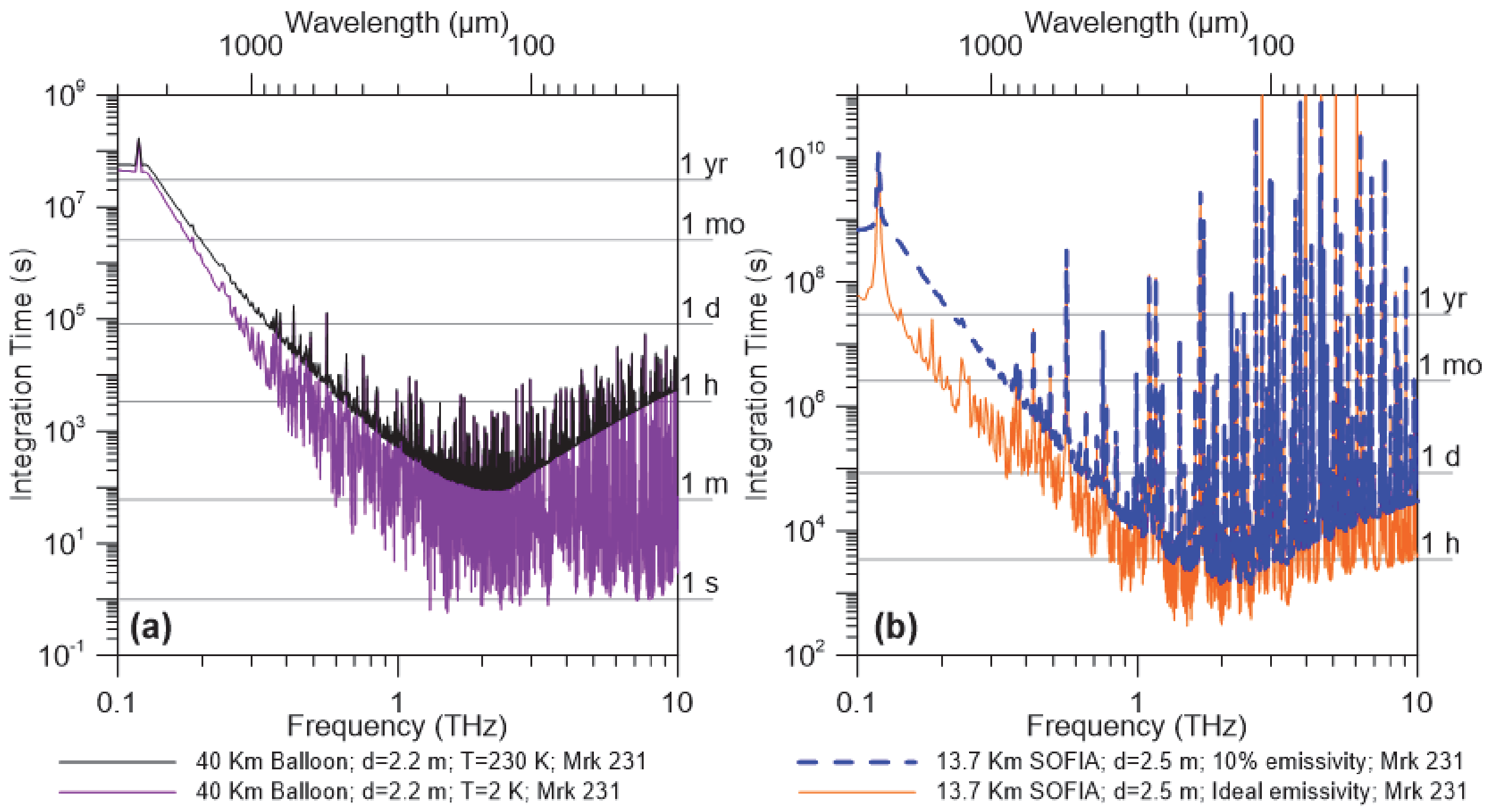

**Figure 4.4:** Effects of mirror and instrument emissivity on integration time. Graph (a) compares cooled and uncooled 2.2 m UCSB balloon systems, graph (b) shows modeled effects of the 10% excess emissivity on SOFIA versus a completely idealized system. Mrk 231 and NGC 958 templates are at z=1.

It is important to keep in mind that we have done the signal to noise analysis and corresponding integration times for the specific case of a dusty galaxy template for concreteness. This is simply one example. We have not introduced the noise of detectors here and thus assume them to be not limiting. The purpose here is to show the fundamental limits with the above assumptions. The same "machinery" can be used to calculate SNR's for specific target SED's and specific missions with known detector response and noise. Also critical is that extraction of data for detecting dusty galaxies would normally not be done on a "wavelength by wavelength" basis but rather via template fitting and optimum extraction algorithms. High resolution spectroscopy (such as R=1000 shown above) allows one to not only "look between atmospheric lines" in many cases to achieve extremely high sensitivity, but also allows for detection of high z emission lines from target galaxies such as C-II. This is an extremely powerful technique for searching for early galaxy formation and the cooling lines likely critical in the formation of structure.

## 5. Effects of Varying Resolution

In the results presented so far, we have used a fractional bandwidth resolution $R$ of 1000, which is useful for many spectrographic observations. An $R$ of 1000 allows adequate resolution for detection of many molecular lines and also allows for very deep low atmospheric backgrounds at high altitudes. The same modeling can be used for various values of $R$, with consequent effects on the integration time. To a first order, the BLING scales as $1/\sqrt{R}$ (Fig. 5.1), the Signal as $1/R$ (Fig. 5.2), and the integration time as $R$ (Fig. 5.3). Additionally, lines in the atmosphere will be smoothed out. Ultimately, the specific nature of an observation will dictate the value of $R$ with consequent influence on system applicability. We show one of our examples below for observing an object like Mrk 231 placed at z=1 from a balloon borne platform for an extreme case of a 10 m diameter system with various R from 3 (photometry) to 1000. Note that the integration time is the time to achieve $SNR = 5$ in each resolution bin, and an optimal search would generally combine bins.

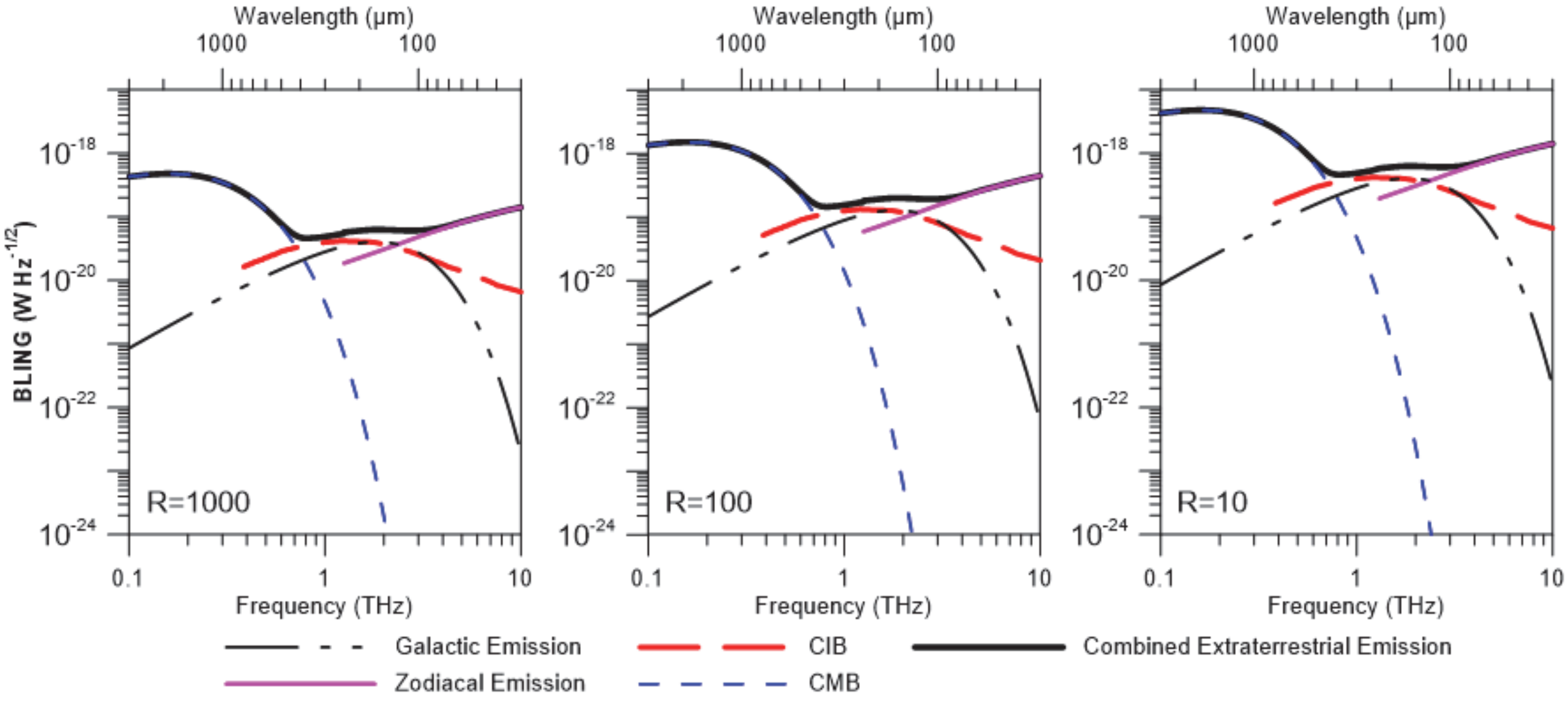

**Figure 5.1:** Extraterrestrial BLING spectra at varied R.

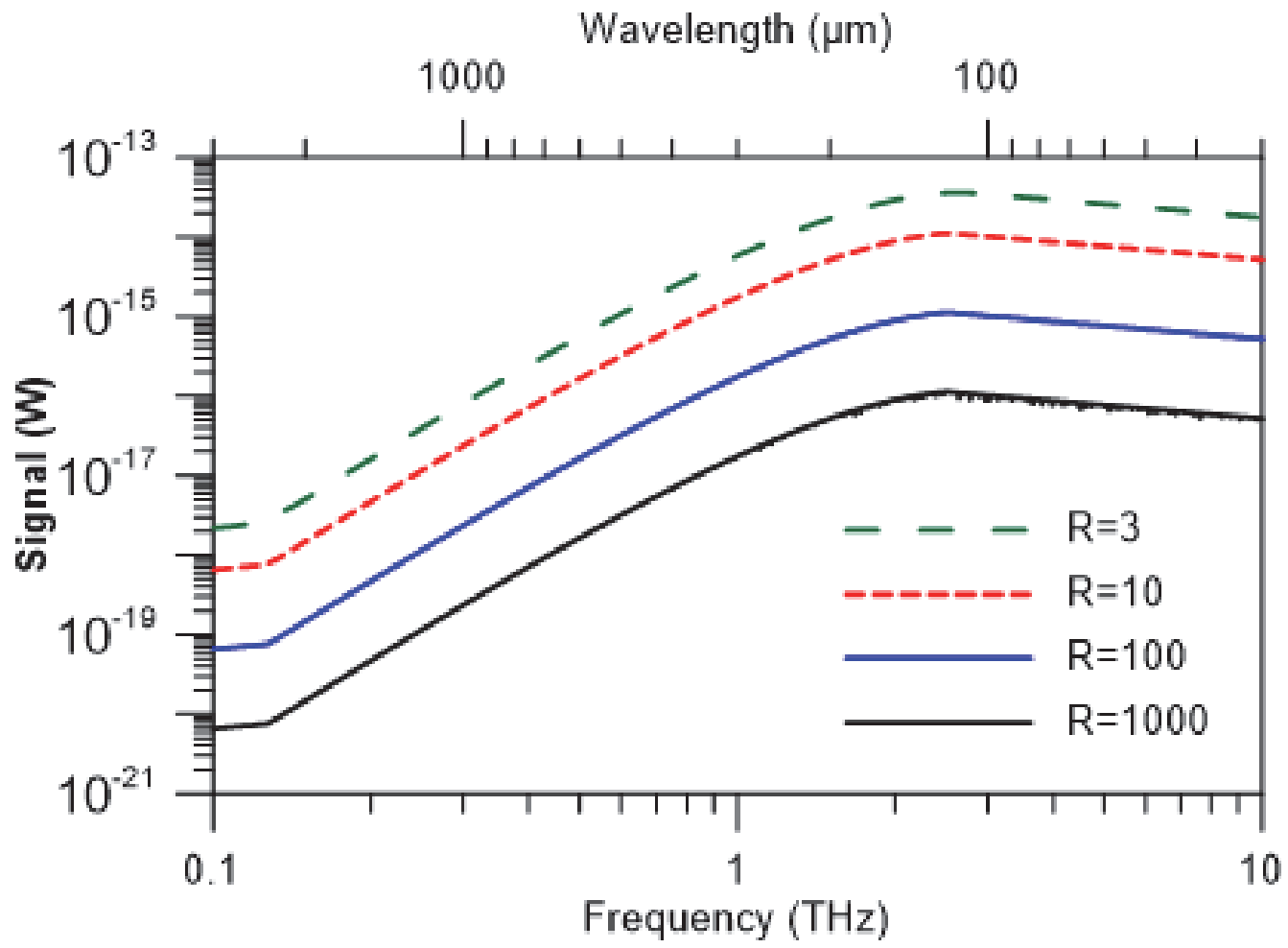

**Figure 5.2**: Signal spectra at varied R of a 40km, 230K, 10m balloon observing a z=1, Mrk 231 galaxy template.

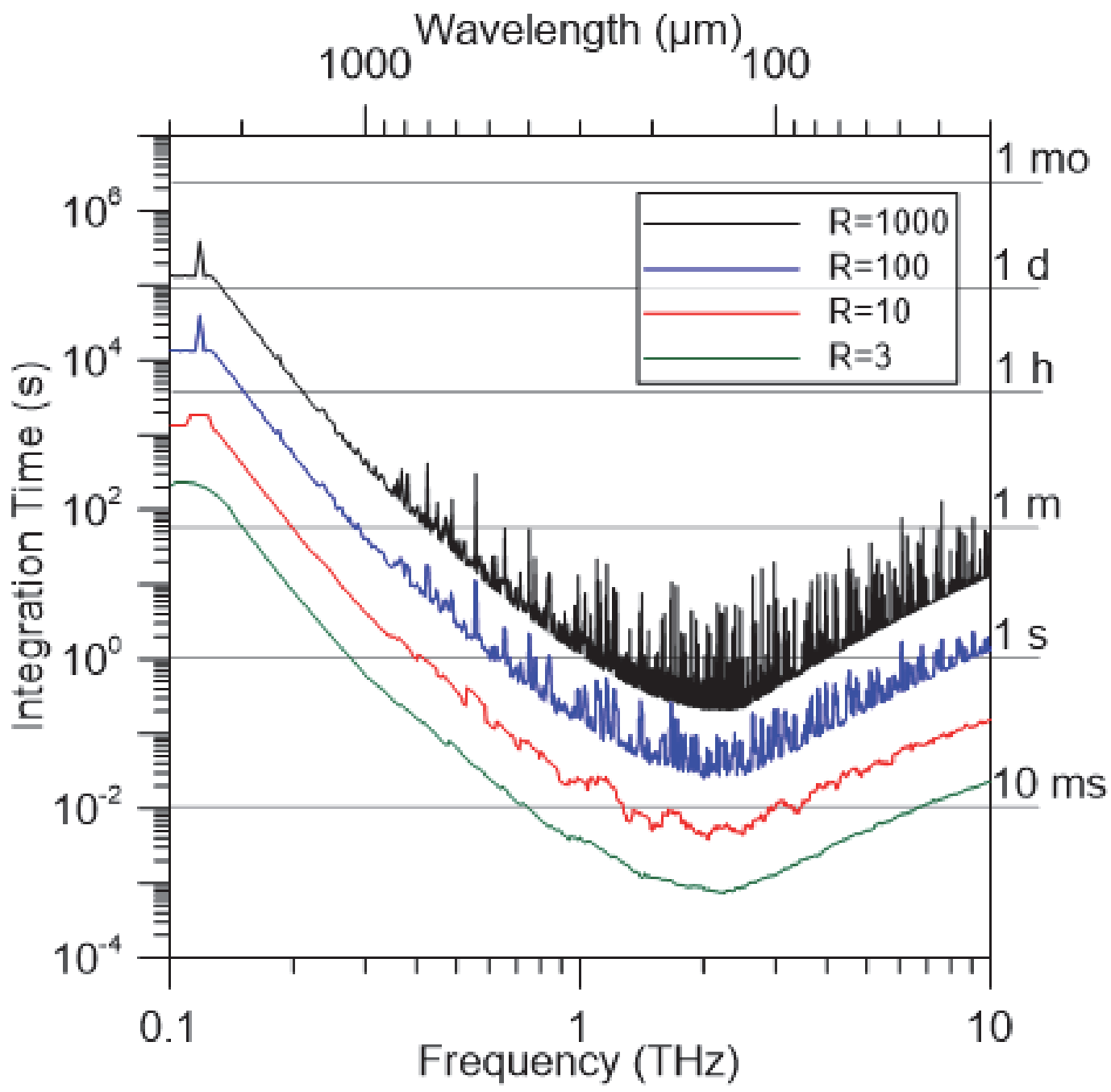

**Figure 5.3**: Integration time at varied R of a 40km, 230K, 10m balloon observing a z=1, Mrk 231 galaxy template.

## 6. Conclusion

We have presented models of fundamental limitations for observations in the far-IR portion of the spectrum and applied this to observations of the continuum emission of high-redshift galaxies. Observing these early galaxies is a significant challenge in modern astrophysics.

Our noise modeling has shown that background photon noise generated by extraterrestrial sources, the CMB, CIB, zodiacal, and galactic emissions is generally small in comparison to atmospheric emission in all cases, even at high altitudes, for frequencies greater than about 0.4THz. For balloon missions with cryogenic optics (a tremendous challenge for larger optics) with temperatures less than 30 K the mirror emission noise is also small relative to atmospheric effects for frequencies greater than ~0.4 THz. However, for balloon cases with warm mirrors, the mirror emission noise term competes with atmospheric noise for dominance. For the best possible ground sites, atmospheric BLING is an order of magnitude or greater than mirror emission from a 230 K mirror above 0.4 THz.

For observing dusty galaxies in the redshift range 1-20, the inverse K-correction has a dramatic impact occurring in frequencies less than 1 THz. Atmospheric transmission to ground sites in the 0.1-10 THz frequency range varies dramatically with PWV, which varies dramatically with day/ night cycles and overall weather, both short and long term. Even ground based sites with the lowest PWV experienced significant signal loss due to atmospheric water content in the far-IR, particularly above 1 THz. Ground sites have reasonable atmospheric transmission in the 0.1-0.3 THz range, with the exception of a few narrow absorption lines, but performance is modest to poor in the 0.5-10 THz range, except in a few narrow windows. In contrast, the low atmospheric water content at high altitudes allows the 40km balloon to have transmission close to unity throughout the entire 0.1-10THz range.

The significantly decreased BLING and much greater atmospheric transmission at high altitudes results in mapping speeds in many areas of the far-IR spectrum several orders of magnitude greater than what can be achieved at the best ground sites, despite the greater aperture size available on the ground.

For frequencies below ~0.4 THz, the integration time required for good ground based sites is less than that of the balloons and space missions owing to the much larger aperture possible (e.g. the proposed 25 m CCAT). Above 0.4 THz (until about 10 THz) ground sites are generally only usable in narrow atmospheric windows. Once in the 1-10THz region, sensitive ground based observations are extremely difficult. Not to be overlooked, however, is that large aperture ground based telescopes give correspondingly higher angular resolution, which is critical in cases of dense fields or searching for fine scale structures. For confusion limited cases angular resolution becomes critical. Thus ground, airborne, and space systems are highly synergistic. A large aperture cold telescope in space, of course, would be the ideal.

While a space mission would be an excellent solution, the high cost and development times place them at a significant practical disadvantage. By the very nature of flight, an aircraft platform like the SOFIA mission still encounters a significant portion of the atmosphere and currently has significant optical emissivity. As a result, modeled integration times for SOFIA are around 3-4 orders of magnitude greater than a 10 meter class balloon payload and about two orders of magnitude greater than a cryogenic 2 m class balloon borne system in the far-IR. These results are significant motivating factors to pursue high altitude balloon borne telescope systems as a method for observing in the far-IR in general, and dusty galaxies in particular. Spectroscopic missions from balloon altitudes where one can observe "between the atmospheric lines" promise exceeding high sensitivity. Finally, we note that our idealized models do not include specific instruments efficiency, but can be easily extended to incorporate them, as proposed telescopes are designed and refined. In a future work, we plan on extending our models to shorter Infrared wavelengths, as well as extending our results to the search for high z emission lines and specific missions.

## Acknowledgements

The authors would like to thank W. T. Reach of the NASA Ames Research Center for generous assistance with the zodiacal emission models.

## References

Becklin, E. E., “Stratospheric Observatory for Infrared Astronomy (SOFIA)”, Proceedings of the ESA Symposium "The Far Infrared and Submillimetre Universe" 15-17 April 1997, Grenoble, France, ESA SP-401 (August 1997) pp. 201-206

Bertoldi, F., Carilli, C.L., Cox, P., et al., 2003, A&A, **406**, L55

Blain, A.W., 1999, MNRAS **309**, 4, 955

Blain, A.W., Smail, I., Ivison, R.J., Kneib, J.-P., Frayer, D.T., 2002,
Ph.R., **369**, 111

Chaplin, Martin. "Water Structure and Science." Water absorption spectrum. May 31, 2011. June 2, 2011. <http://www.lsbu.ac.uk/water/vibrat.html#blue>

Committee for a Decadal Survey of Astronomy; Astrophysics; National Research Council, *New Worlds, New Horizons in Astronomy and Astrophysics*. The National Academies Press: 2010. <http://sites.nationalacademies.org/bpa/bpa_049810>

Finkbeiner, D. P., Davis, M., & Schlegel, D. J. 1999, ApJ **524**, 867

Franceschini, A., Aussel, H., Cesarsky, C. J., Elbaz, D., Fadda, D.,2001, A&A **378**, 1

Franceschini, A., Rodighiero, G., Vaccari, M., 2008, A&A **487**, 3

Hauser, M. G., Dwek, E., “The Cosmic Infrared Background: Measurements and Implications,” Annual Review of Astronomy and Astrophysics, 2001, **39**, 249-307

Herter, T., Radford, S. "Atacama PWV Distribution." CCAT Tech Memos TH-05-003. 2005. Cornell University. June 2, 2011 <http://wiki.astro.cornell.edu/twiki/bin/view.pl/CCAT/CCAT_Memos>.

Hogg, D. W. 1999 arXiv:astro-ph/9905116

Kelsall, T. et al. 1998, ApJ, 508, 44

G. Lagache et al. "Modeling infrared galaxy evolution using a phenomenological approach", MNARS **338**, 3, 555–571
B. Magnelli et al. "Release of the deepest Herschel -PACS far-infrared survey: number counts and infrared luminosity functions from combined PEP/GOODS-H observations", Astronomy & Astrophysics, in press 2013

Marvil, J. et al., 2005, New Astronomy **11**, 218-225

Nakagawa, T. et al. "SPICA Mission: future Japanese infrared astronomical mission."SPICA Mission: future Japanese infrared astronomical mission. 1998. June 2, 2011 <http://www.ir.isas.jaxa.jp/SPICA/h2l2_spie/h2l2.html>.

NASA/IPAC Extragalactic Database, Spectral Energy Distribution for ARP 220 and NGC 958. Retreived February 23, 2013. < http://ned.ipac.caltech.edu/>

de Oliveira-Costa, A., Tegmark, M., Gaensler, B. M., Jonas, J., Landecker, T. L., & Reich, P. 2008, MNRAS **388**, 247

Pardo, J. R., Cernicharo, J., Serabyn, E., "Atmospheric Transmission at Microwaves (ATM): An Improved Model for mm/submm applications", 2001, IEEE Trans. on Antennas and Propagation, 49/12, 1683-1694

Pascale, E., Ade, P. A. R., Bock, J. J., et al. 2008, ApJ**681**, 400

PcModWin5 v2r0, Ontar Corporation, North Andover, MA, United States.

Radford, S. "Cerro Chajnantor Water Vapor and Atmospheric Transmission", CCAT Tech Memo 112, Dec 2012

Soifer, B.T., Neugebauer, G., Houck, J.R., 1987 Ann. Rev. Astron. Astrophys., **25**, 187S

Wright, E.L.,2003, Carnegie Observatories Astrophysics Series **2**.
arXiv:astro-ph/0305591

Xu, J., Lange, A.E., Bock, J.J., "Far-Infrared Emissivity Measurements of Reflective Surfaces." Submillimetre and Far-Infrared Space Instrumentation, Proceedings of the 30th ESLAB Symposium (ESTEC, Noordwijk, The Netherlands, 1996), pp. 69–72

Yang, H. et al., 2010, PASP **122**, 490-494